\documentstyle{article}
\newcommand{\be}{\begin{eqnarray}}
\newcommand{\ee}{\end{eqnarray}}

\newcommand{\ben}{\begin{enumerate}}
\newcommand{\een}{\end{enumerate}}
\newcommand{\bs}{\bigskip}

\begin{document}
\rightline{McGill/96-28}
\input epsf.tex
\begin{center}
{\huge\bf Gauge cancellation for electroweak processes and
the Gervais-Neveu gauge}
\end{center}
\bs
\begin{center}
{Y.J. Feng$^*$ and C.S. Lam$^{\dag}$}\\
\bs
{\it Department of Physics, McGill University,\\
3600 University St., Montreal, P.Q., Canada H3A 2T8}
\end{center}

\begin{abstract}
A graphical method is developed to study the total or partial
cancellation of gauge-dependent (divergence) terms in electroweak
theory. The method is used to work out rules in the Gervais-Neveu
gauge, whose triple-gauge vertex contains three terms rather
than the usual six. This and other features of the gauge lead to
an enormous saving of algebraic labor in the actual computations,
as will be illustrated explicitly in the
tree process $W^++W^-\to \gamma+\gamma$.
\end{abstract}

\section{Introduction}
Standard-Model calculations are beleaguered with gauge problems.
The algebra is made unnecessarily complicated because individual Feynman
diagrams depend on the gauge chosen, although these gauge-dependent terms
must all get cancelled out in the sum. While there is no known way 
to eliminate this complication, a clever choice of gauge can greatly 
simplify the ensuing calculations. The purpose of this paper
is to discuss ways of doing that in the electroweak theory (EW).

Gauge choices can be made on every Feynman-diagram component connected
to a gauge particle. Different gauges differ by
a divergence factor, {\it i.e.,} terms proportional to 
the momentum of the gauge particle.
Gauge-dependent terms are made out of these divergence factors;
a study of their cancellation is a study of how such
divergences from the various diagrams combine to annihilate one another.
The purpose of the previous paper \cite{YJ} and the present one
is to develop a simple technique to study this general problem,
and hence a tool to help us decide on the best gauge to use.
The simplicity results from our systematic use of the graphical language,
thus avoiding the very messy algebra otherwise needed. For QCD a 
detailed discussion can be found in Refs.~[1] and \cite{new}.

There are three different components of a Feynman diagram where
gauge choices can be made: the gauge propagator, the external
gauge-particle wave function, and vertices involving a gauge
particle. For QCD the gauge propagator is usually
chosen to be in the Feynman gauge. For
EW it is either the Feynman gauge or the unitary
gauge.
As far as external gauge-particle wave function
is concerned, enormous simplification can be obtained using the
spinor helicity method, originally developed to be used with
tree diagrams \cite{RG}, but with the introduction of the electric-circuit 
technique \cite{CS}, super-string
\cite{ZB} 
or first-quantized \cite{MS}
 formalism, the method can equally be
used to compute loop diagrams. The remaining
gauge-dependent components are the vertices. For QCD \cite{YJ}, 
simplifications can often be obtained by avoiding using 
the ordinary three-gluon vertices containing
six terms. In tree diagrams, the simplest vertices available
are those in the Gervais-Neveu (GN) gauge \cite{JL}, where the triple gluon
vertex contains only three, rather than six, terms. 
For one-loop one-particle irreducible
$n$-gluon diagrams, in some sense the background-field method 
(BFM) \cite{BS} offers
the greatest economy. For other processes, or two and more loops,
it is not known what the best gauge would be, but it will generally
not be the ordinary nor the BFM gauge, simply because a new gauge can be
constructed for two-loop gluon self-energy diagrams
which gives rise to simpler results than either of those two gauges \cite{new}.
 
For electroweak theory, like in QCD, graphical methods 
can be used to fix gauges. In particular, one can derive the
rules to be used in the GN gauge which
we will discuss in Sec.~3, and
we shall see that 
its vertices offer
enormous simplifications. A comparison for the savings will be illustrated
by using the tree process $W^++W^-\to\gamma+\gamma$.
 
This method can also be used to determine how computations can be
carried out in other gauges. When applied to obtaining the BFM gauge \cite{XL}
for one-loop electroweak calculations, we reproduce
the results of the pinching technique \cite{JP}.

\section{Divergence relations and gauge cancellations}

The discussion of gauge transformation and gauge cancellation
is necessarily more complicated than the corresponding case
in QCD, owing to the following two facts.  First the gauge particles ($g$)
can now be either $W^\pm, Z$, or $\gamma$, with different
masses and different number of polarizations. Secondly, because of
spontaneous symmetry breaking, the nonabelian
group factor can no longer be isolated.
This increases the number of Feynman rules as shown in App.~A.
Other than these two differences and the associated complications, the
discussions are  completely parallel to those in QCD \cite{YJ}. 

Throughout this paper, we choose Feynman gauge for all the internal propagators,
as we did in QCD \cite{YJ}. 
We shall focus in the text 
new aspects appearing in the electroweak theory, leaving other details
to the graphical formulas in App.~B.
As before, we first study the divergences of all the vertices that contain at 
least one gauge boson ($g$), and then how the various terms so generated
are cancelled or partially cancelled.

First let us review the Ward-Takahashi identity in QED. 
This can be found in many text books \cite{Sterman} on quantum field 
theory.
Graphically, the identity is shown in Figs.~1(a)--1(c). 
{\it A cross on a gauge boson (wavy) line always
 denotes the divergence, which is just a 
factor $p_{\alpha}$ for the gauge boson with outgoing momentum $p$ and
Lorentz index ${\alpha}$}. A dotted line is a ghost line ($G$). When drawn
tangential to a propagator, it simply injects the right amount of momentum
into the propagator without changing anything else. This happens in
diagrams (b), (c), (e), (f), (k), and (l), which we shall call the
{\it sliding} diagrams. Here and after, each diagram in an identity should
be understood as part of a large Feynman diagram. The small dot at the end of
a line indicates that its propagator should be included. 
\begin{figure}
\vskip -0 cm
\centerline{\epsfxsize 4.7 truein \epsfbox {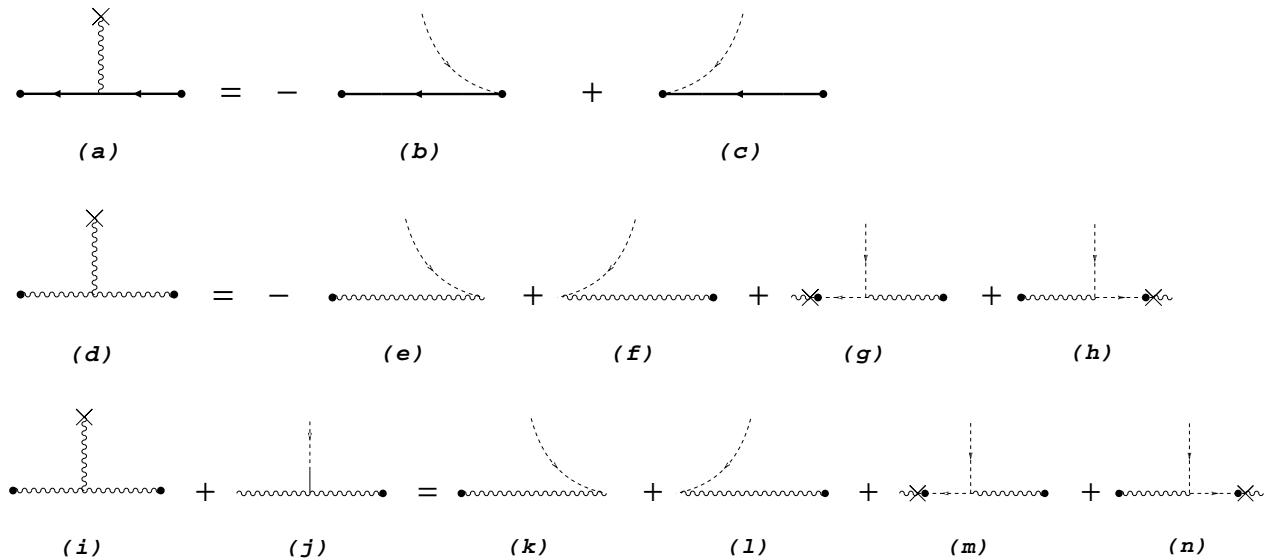}}
\nobreak
\vskip -7.5cm\nobreak
\vskip .1cm
\caption{Divergence relations for QED, QCD, and EW.
All momenta are outgoing; all particles except the ghost are also outgoing.}
\end{figure}

The corresponding identity in QCD is shown in Figs.~1(d)--1(h). 
The complication arising 
from color source diffusion of a nonabelian theory 
is reflected in the {\it propagating} diagrams 
Figs.~1(g) and 1(h). 
Here a cross at the end of a ghost line represents a cross
on the gauge-boson line it is connected to.  
A propagating cross always
drags behind it a ghost line, which is a metamorphosis of
the gauge-boson line. To distinguish this from an internal ghost line
appearing in a loop, this will be called a {\it wandering ghost line}.

This identity in the case of EW theory (Figs.~1(i)--1(n)) 
is even more complicated on account of the 
spontaneous symmetry breaking. This leads to the appearance of the 
{\it compensation} diagram (Fig.~1(j)), so called because 
it compensates the mass difference of those two sliding diagrams.
To illustrate this, we write the divergence of a $WW\gamma$ vertex.
\begin{eqnarray}
&&(p_2)^{\beta}T_{\alpha\beta\delta}(p_1,p_2,p_3)+(-i M_W)i e g_{\alpha\delta} M_W  \nonumber\\
&=&eg_{\alpha\delta}[-p_3^2+M_W^2]  
+eg_{\alpha\delta}[p_1^2] 
+e(p_1)_{\alpha}[-(p_1)_{\delta}] 
+e (p_3)_{\delta}[(p_3)_{\alpha}]\ . 
\end{eqnarray}
These terms are shown respectively as Figs.~1(i), 1(j), $\cdots$, 1(n).
A solid line represents a scalar field $S=\{\phi,H\}$, where $\phi$ are the
Goldstone fields and $H$ is the physical Higgs particle.
When a ghost line is joined to a solid line $\phi$ 
 as shown in Fig.~1(j),
 we assign to the two-point vertex a factor $-i M_W$.

Note that we have
redefined the sign of the sliding diagram in EW compared to those used
in QED and QCD \cite{YJ}. This simplifies later discussions 
as will be seen in App.~B.

There are many divergence relations shown in App.~B. Their general features
needed for later discussions will be summarized below.

A divergence relation is an identity with a $g$-line crossed in
one or two of its vertices. Those diagrams with a cross will be called
{\it divergence diagrams}.
If the vertex being crossed
does not contain ghost line itself, then there is only one divergence diagram
in an identity. If the crossed vertex contains ghost line, then there will be
two divergence diagrams (see for example Fig.~18). 

We first look at the 
the divergence relations in App.~B 
with only one divergence diagram. They will be summarized below.
By definition, all diagrams carry a coefficient $+1$, so it is important
to know which of them appear on the lefthand side and which of them
appear on the righhand side of the relations. 

First the lefthand side of the divergence relations.

Replace the $g$-line with a cross
by its corresponding  $\phi$-line.
If this new vertex is a legitimate Feynman-rule vertex, 
then the lefthand side of the relation consists of the divergence diagram
as well as a compensation diagram.
See for example Figs.~1(i) and 1(j). 

Now imagine we remove the crossed
$g$-line from the divergence diagram. If the new vertex is still
a legitimate Feynman-rule vertex, then the lefthand side consists of 
the divergence diagram as well as sliding diagrams.
They are constructed by changing
the crossed $g$-line into a wandering ghost line. See for
example Fig.~16.

Note that only one of these two possibilities occurs in App.~B. However,
in the GN gauge to be discussed in Sec.~3, both of these possibilities
can occur simultaneously. See Fig.~43.

Now the righthand side of the divergence relations.

Propagating diagrams,
sliding diagrams, and compensation diagrams appear on the righthand side 
according to the following rules.

First, if the vertex being crossed is a three point vertex, and if those two 
uncrossed lines are both $g$-lines, fermion lines, $S$-lines, or ghost lines,
then there are sliding diagrams on the righthand side. For the first three 
cases, there are two sliding diagrams each with a propagator being 
cancelled (see for example Fig.~1(k) and 1(l)). For ghost vertex, 
there is only one sliding diagram 
with the propagator of the outgoing ghost line being cancelled (see Fig.~19). 

Secondly, change the crossed $g$-line
into an incoming ghost line and any one
of the uncrossed $g$-lines into an outgoing ghost 
line. If the new vertex is legitimate,  then there is a propagating
diagram constructed with this change as shown in Figs.~1(m) and 1(n). 

Thirdly, there may
be compensation diagram(s) as well, as long as we can legitimately 
change one of the 
$\phi$-lines in the
divergence diagram into an outgoing ghost line, and simultaneously
the crossed $g$-line into an 
incoming ghost line (see for example Fig.~12(e)). 

Now we are going to look at the divergence relations containing 
two divergence diagrams on the lefthand side. The diagrams in these
identities carry both signs.

In the Feynman gauge,
this occurs only in connection with the $GGg$ vertex -- the divergence
diagram obtained by interchanging the crossed $g$-line with the incoming
ghost line has an opposite sign. This can be traced back to 
the minus sign of the ghost loop. 

For either of
these two divergence diagrams, we can use the rules discussed before 
to determine the rest of the diagrams in the relations.
See Figs.~19--21.

In the divergence relations, we have included all the possibilities of
a cross being placed on any $g$-line of a vertex. To complete the list of
all the identities, we need to consider relations without crosses. 
We shall call those {\it cancellation relations}. They
involve diagrams of
following two kinds: 1) An incoming ghost line $G$ is joined to a
$\phi$-line
of a vertex; and 2)  
A wandering ghost line is tangential to a propagator. Every term below
once again has a coefficient $+1$.

In the first case, if the vertex obtained by replacing the incoming
ghost line by a crossed $g$-line is legitimate, then the identity
has been considered before and is a divergence relation. 
Otherwise, terms with the $G-\phi$ line
replaced by tangential ghost lines are also present on the lefthand
side of the cancellation relation in all possible ways. 
The righthand side of the identity is zero. See Fig.~24 for an example.

In the second case, we must first decide whether the sliding diagram
is already involved in a divergence relation, or an identity under case 1.
If not, sum up all the possible sliding diagrams to get zero as the identity.
See Fig.~30 for an example.

These are all the identities in Feynman gauge that we need to know.

\subsection{U(1) gauge symmetry}

As an application of these identities, we shall demonstrate explicitly
how the remaining $U(1)$ gauge symmetry is preserved in the diagrammatic
language.
This means that if we introduce a gauge transformation to one of 
the external photons,
\begin{eqnarray}
\epsilon^{\mu}(p)\to\epsilon^{\mu}(p)+\lambda p^{\mu} \ ,
\end{eqnarray}
with $\lambda$ being an arbitrary parameter, then
the contribution coming from this
extra divergence term must sum up to zero, provided all the other external 
gauge bosons are transversely polarized. 

The verification of this is very similar to the case of {\bf R1} 
considered in QCD \cite{YJ}.
Roughly speaking it is the following. 
Using the divergence relation on the gauge
vertex involving the external photon, propagating, sliding, and compensation 
diagrams
are generated. The crosses in the propagating diagrams can be used
in other divergence relations to produce other propagating, sliding,
and compensation diagrams, and so on. Whatever sliding or compensation
diagrams needed on the lefthand side of a divergence relation will 
be automatically produced from the last divergence relation. Continuing thus,
finally many diagrams are generated for the divergence of the amplitude.
The cross is either absent in a diagram, or else it appears in an
external $g$-line, in which case the corresponding diagram vanishes provided
that gauge particle is transversely polarized. The remaining diagrams without
crosses sum up to zero by using the cancellation relations. When divergence
relations involving two divergence diagrams are used, the argument becomes
more complicated because of the presence of the minus signs, but these signs
can always be traced back to the signs of the ghost loops so everything
will again work out properly. Consequently, just as in QCD \cite{YJ}, we have

\begin{itemize}
\item
 {\bf R1}: The sum of all on-shelled or crossed diagrams with one cross
 on one of the
external photon lines is zero, if all the internal propagators are taken in 
the Feynman gauge. This is just the remaining U(1) gauge symmetry.
\end{itemize}

\subsection{Equivalence theorem}
Another application of these graphical identities is the verification
of the equivalence theorem \cite{JM}.
After spontaneous symmetry breaking, the W and Z bosons obtain masses so
their external wavefunctions are no longer gauge invariant.
From the point of view of the identities in App.~B, 
this lack of invariance is brought about because the corresponding
compensation diagram on the left of the divergence relation is now absent.

Since all 
the longitudinally polarized external 
wave-function of a massive boson is
\begin{eqnarray}
{1\over M_W} (p,0,0,p_0)\ .
\end{eqnarray}
Taking the high energy limit 
\begin{eqnarray}
E_j\sim k_j\gg M_W,
\end{eqnarray}
the external wave-function of a W 
particle can be represented by a cross. This would have produced a zero
result at the end if the compensation diagram were present on the lefthand
side of the divergence relation. Consequently, longitudinally polarized
W boson is equivalent to the corresponding $\phi$, which is the content
of the equivalence theorem \cite{JM}.

\subsection{Pinching technique and Background Field Method}
In the previous paper \cite{YJ} we have shown the affinity between our
graphical language and the pinching technique \cite{JP} in the case
of QCD. The latter can be obtained from the former by changing the
vertex to the BFM gauge in the one-loop situation \cite{YJ,XL}. Now that we
have established similar graphical rules for EW, the same connection
can be shown in essentially the same way.

\section{Gervais-Neveu Gauge}
The Gervais-Neveu (GN) gauge was first discussed in Ref.\cite{JL} for QCD, 
and for a toy 
model of massive non-abelian Yang-Mills fields. It was found out later that 
superstring selects the GN gauge for its QCD tree-level calculations\cite{ZB}.
This is gratifying because in some sense the GN gauge is the simplest 
gauge to use at the tree level,  both for
QCD and EW. As a matter of fact, this gauge
simplifies computations in the loop levels as well. 
In this section we shall use the graphical identities to work out
the vertices and their Feynman rules in the GN gauge, and we shall discuss
a simple example illustrating the amount of labour that can be saved by 
its use.

To obtain the GN gauge in EW we adopt a procedure very similar to the one used
in QCD where the BFM gauge was derived from the Feynman gauge \cite{YJ}. 
To avoid
a paper with excessive length, we refer the readers to Ref.~[1] for
most of the details. Nevertheless, 
we shall summarize the procedure below, with special
emphasizes on points that are peculiar to the EW theory. 

The Feynman rule for vertices in the GN gauge is not the same
as those in the Feynman gauge. In addition, there are new vertices
present in the GN gauge. Both of these are summarized in Tables I and II.

The procedure to obtain them can be outlined as follows:

\begin{enumerate}
\item 
By using momentum conservation, the six terms in the usual $3g$ vertices
can be rewritten as the sum of three divergence terms (Figs.~2(c), (d), (e)),
and a new GN vertex (Fig.~2(b)), as shown in the formula below:
\begin{eqnarray}
&&\lambda \left(g_{\alpha\beta}(p_1-p_2)_{\delta}+
               g_{\beta\delta}(p_2-p_3)_{\alpha}+
               g_{\delta\alpha}(p_3-p_1)_{\beta}\right) \nonumber\\
&=&2\lambda \left(g_{\alpha\beta}(p_1)_{\delta}+
                  g_{\beta\delta}(p_2)_{\alpha}+
                  g_{\delta\alpha}(p_3)_{\beta}\right) 
+ \lambda g_{\alpha\beta} (p_3)_{\delta} 
+ \lambda g_{\beta\delta}(p_1)_{\alpha} 
+ \lambda g_{\delta\alpha}(p_2)_{\beta}\ , 
\end{eqnarray}
where $\lambda=e$ for $W^+W^-\gamma$ coupling, and $\lambda=g\cos\theta_W$ 
for $W^+W^-Z$
coupling.

\item Results obtained in Secs.~2.1 and 2.2 will be used to show that
the three divergence terms combine to cancel one another partially.
The remaining terms will combine with other existing vertices to
obtain the new vertices shown in Tables I and II.
\end{enumerate}

\begin{figure}
\vskip -0 cm
\centerline{\epsfxsize 4.7 truein \epsfbox {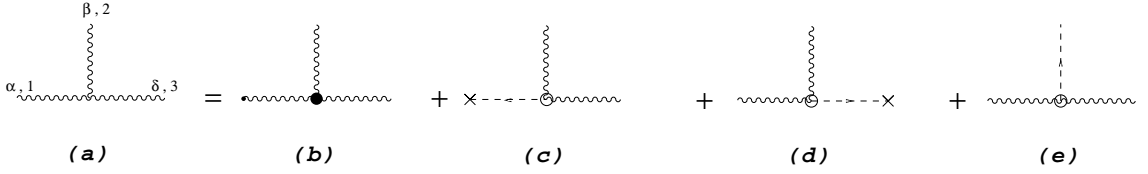}}
\nobreak
\vskip -14.2cm\nobreak
\vskip .1cm
\caption{Relation between the $3g$ vertex in Feynman gauge and in GN gauge.
The GN vertex is denoted by a big dot.}
\end{figure}

\begin{figure}
\vskip -.2 cm
\centerline{\epsfxsize 4.7 truein \epsfbox {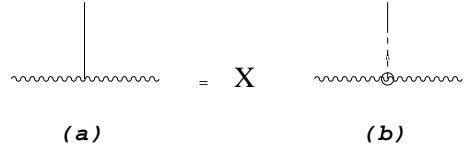}}
\nobreak
\vskip -14.5cm\nobreak
\vskip .1cm
\caption{An alternative way to write the $gg\phi$ vertex.}
\end{figure}

We shall now discuss in more detail how step 2 above is carried out.

\begin{itemize}
\item[2A.] Consider all the diagrams containing 2(c), (d), or (e).
Imagine for the moment that the incoming
ghost line containing the cross is cut open.
Apply arguments similar to those used in Secs.~2.1 and 2.2, all diagrams should
sum up to be zero if the presence of the 
external outgoing ghost on the other side of the cut, as well as the
subtlety to be mentioned below, are ignored.
 
The subtlety is the following.
In the EW theory
there is a distinction between a neutral particle ($\gamma, Z$)
and a charged particle ($W^\pm$), in that the compensation diagram is
required on the lefthand side of the divergence relations for the latter.
The compensation diagram required in the present context must have
a ghost line connected to the other side because this is what the crosses
in 2(c), (d), and (e) are connected to. For that reason it is convenient
to rewrite the $gg\phi$ vertex in the form of Fig.~3, where
$X=1$ for $\gamma(Z)W^-\phi^+$ 
coupling and $X=-1$ for  $\gamma(Z)W^+\phi^-$. 
If $X=1$, the compensation diagram comes in correctly for the divergence
relation to be used. 
If $X=-1=-2+1$, then on top of the compensation diagram, the coefficient
of the $gg\phi$
vertex is now doubled in the GN gauge.

In addition, the actual presence of the outgoing ghost must be taken into
account. This gives rise to some `leftover terms' and causes an incomplete
cancellation. It is from these leftover terms, combined with other existing
vertices, that new vertices and/or new Feynman rules are obtained.

\item[2B.] The Feynman rule for the $GGg$ vertex in the GN gauge
is obtained from Fig.~4, in which diagram (a) is a leftover term (see 2A).
Depending on the sign of the ordinary ghost vertex, the corresponding GN ghost 
vertex can be either the `a-type': $(p_1-p_3)_{\beta}$, or the `b-type': 
$-(p_3+p_1)_{\beta}=(p_2)_{\beta}$.
The exact result is shown in Table II. 

\item[2C.] There is a new $GGgg$ vertex created in the GN gauge,
obtained from one or two leftover terms as shown in Figs.~5 and 6.
Which is which depends on the details and the results are shown in
Table II.

\item[2D.] These GN vertices are obtained diagrammatically by using
Fig.~2 for one of the possibly many $3g$ vertices in the diagram.
Once this calculation is completed we will proceed to make the same
decomposition for other $3g$ vertices, one after another. At that point,
since there may already be some GN vertices present, in principle
new identities are required and still newer vertices can be created.
These identities are shown in App.~C. The net result is that only one more
new vertex is created this way in the GN gauge, which is discussed in
the item below.

\item[2E.] The $4g$ vertex in the GN gauge shown in Fig.~7(d) is
a combination of the usual $4g$ vertex and some leftover terms.
See Table II for the final result.

\end{itemize}

\begin{figure}
\vskip -1 cm
\centerline{\epsfxsize 4.7 truein \epsfbox {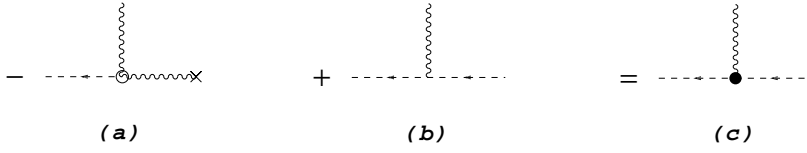}}
\nobreak
\vskip -13.5cm\nobreak
\vskip .1cm
\caption{The combination that generates the $gGG$ vertex in the GN gauge.}
\end{figure}

\begin{figure}
\vskip -.5 cm
\centerline{\epsfxsize 4.7 truein \epsfbox {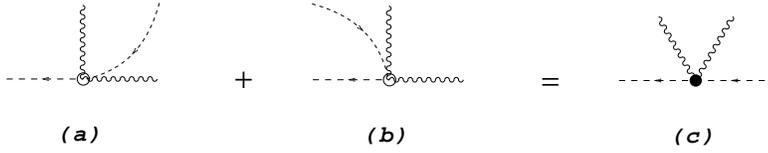}}
\nobreak
\vskip -13.5cm\nobreak
\vskip .1cm
\caption{The generation of one kind of $GGgg$ vertices in the GN gauge.}
\end{figure}

\begin{figure}
\vskip -0.5 cm
\centerline{\epsfxsize 4.7 truein \epsfbox {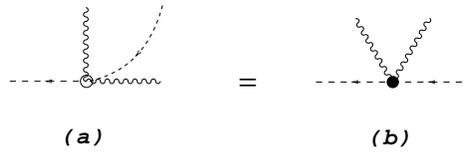}}
\nobreak
\vskip -13.5cm\nobreak
\vskip .1cm
\caption{The generation of the other kind of $GGgg$ vertices in GN gauge.}
\end{figure}

\begin{figure}
\vskip -0 cm
\centerline{\epsfxsize 4.7 truein \epsfbox {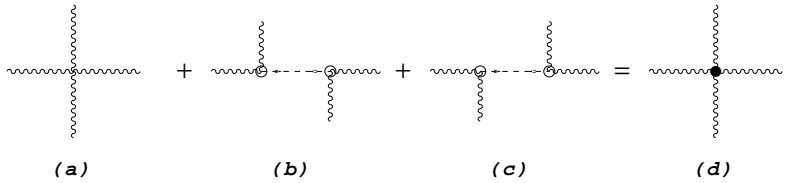}}
\nobreak
\vskip -13.5cm\nobreak
\vskip .1cm
\caption{The generation of $4g$ vertex in GN gauge.}
\end{figure}

The final vertex factors for GN gauge is given in the following tables. 
Here we just list the new vertices and the ones 
that are different from the ones in
Feynman gauge.

\begin{center}
\begin{tabular}{|l|l|l|l|}   \hline
\multicolumn{4}{|c|}{Table for three-point vertices}\\ \cline{1-4}\hline
1 ($\alpha$)  &2 ($\beta$)         &3 ($\delta$)& vertex factor \\  \hline
$\phi^+$      &      $\gamma, Z$    &  $W^-$    &     0   \\   \hline
$W^+$         &      $\gamma$       &  $W^-$    &
 $2 e(g^{\alpha\beta} p_1^{\delta}+g^{\beta\delta}p_2^{
                             \alpha}+g^{\delta\alpha}p_3^{\beta})$   \\ \hline
$W^+$         &      $Z$      &  $W^-$    & $2 g \cos\theta_W(g^{\alpha\beta} p_1^{\delta}+g^{  \beta\delta}
p_2^{\alpha}+g^{\delta\alpha}p_3^{\beta})$   \\ \hline
$W^+$     &       $\gamma$       &  $\phi^-$   &  $2 ieM_W g^{\beta\delta}$   \\ \hline
$W^+$     &     $\phi_z$  &  $W^-$      &  $ig\cos{\theta}_W  M_z g^{\alpha\delta}$   \\ \hline
$W^+$     &       Z       &  $\phi^-$   &  $-2ig\sin^2\theta_WM_zg^{\alpha\beta}$  \\ \hline
$\bar{c}^-$         &  $W^+$    &   $c_{\gamma, z}$& $e(p_1-p_3)_{\beta},\ 
g \cos\theta_W (p_1-p_3)_{\beta} $  \\ \hline
$\bar{c}^+$         &    $\gamma, Z$      & $c^-$ & $e(p_1-p_3)_{\beta},\ 
g\cos\theta_W (p_1-p_3)_{\beta}$ \\ \hline
$\bar{c}_{\gamma, z}$  &  $W^-$    & $W^+$ & $e(p_1-p_3)_{\beta},\ 
g\cos\theta_W (p_1-p_3)_{\beta}$  \\ \hline
$\bar{c}^+$         &  $W^-$    & $c_{\gamma, z}$ & $-e(p_1+p_3)_{\beta},\ 
-g\cos\theta_W (p_1+p_3)_{\beta}$ \\ \hline
$\bar{c}^-$         &  $\gamma, Z$      & $c^+$ & $-e(p_1+p_3)_{\beta},\ 
-g\cos\theta_W (p_1+p_3)_{\beta}$ \\ \hline
$\bar{c}_{\gamma, z}$  &  $W^+$    & $c^-$ & $-e(p_1+p_3)_{\beta},\ 
-g\cos\theta_W (p_1+p_3)_{\beta}$ \\ \hline
\end{tabular}
\end{center}
\begin{center} {Table I}\end{center}

\begin{center}
\begin{tabular}{|l|l|l|l|l|}   \hline
\multicolumn{5}{|c|}{Table for four-point vertices}\\ \cline{1-4}\hline
1 ($\alpha)$&2 ($\beta$)   &3 ($\delta$) &4 ($\rho$) & vertex factor \\ \hline
$W^+$     &   $W^+$  &  $W^-$    &  $W^-$    & $g^2 (2 g^{\alpha\beta}g^{\delta\rho})$ \\ \hline
$W^+$     &    $\gamma$     &   $\gamma$       &   $W^-$   & $e^2(-2 g^{\alpha\rho}g^{\beta\delta}
+2 g^{\alpha\delta}g^{\beta\rho}+2g^{\alpha\beta}g^{\rho\delta})$ \\ \hline
$W^+$     &    Z     &   Z       &   $W^-$   & $g^2 \cos^2\theta_W(-2 g^{\alpha\rho}
   g^{\beta\delta}+2 g^{\alpha\delta}g^{\beta\rho}+2g^{\alpha\beta}g^{\rho\delta})$ \\ \hline
$W^+$     &    $\gamma$     &   Z       &   $W^-$   & $e g \cos\theta_W(-2 g^{\alpha\rho}
   g^{\beta\delta}+2 g^{\alpha\delta}g^{\beta\rho}+2g^{\alpha\beta}g^{\rho\delta})$ \\ \hline
$\bar{c}^+$        &  $W^-$    &   $W^-$  &    $c^+$   &  $-2 g^2 g^{\beta\delta}$   \\ \hline
$\bar{c}^-$        &  $W^+$    &   $W^+$  &    $c^-$   &  $2 g^2 g^{\beta\delta}$   \\ \hline
$\bar{c}^-$        &  $W^-$    &   $W^+$  &    $c^+$   &  $-g^2 g^{\beta\delta}$     \\ \hline
$\bar{c}^+$        &  $W^+$    &   $W^-$  &    $c^-$   &  $g^2 g^{\beta\delta}$     \\ \hline
$\bar{c}_{\gamma}$ &  $W^+$    &     $\gamma$    &    $c^-$   &  $-e^2 g^{\beta\delta}$     \\ \hline
$\bar{c}_{\gamma}$ &  $W^-$    &     $\gamma$    &    $c^+$   &  $e^2 g^{\beta\delta}$     \\ \hline
$\bar{c}^-$        &  $W^+$    &     $\gamma$    &     $c_{\gamma}$      &  $-e^2 g^{\beta\delta}$     \\ \hline
$\bar{c}^+$        &  $W^-$    &     $\gamma$    &     $c_{\gamma}$      &  $e^2 g^{\beta\delta}$     \\ \hline
$\bar{c}^+$        &   $\gamma$       &     $\gamma$    &    $c^-$   &  $-2 e^2 g^{\beta\delta}$   \\ \hline
$\bar{c}^-$        &   $\gamma$       &     $\gamma$    &    $c^+$   &  $2 e^2 g^{\beta\delta}$   \\ \hline
$\bar{c}_z$        &  $W^+$    &     Z    &    $c^-$   &  $-g^2\cos^2\theta_W g^{\beta\delta}$     
                                                                              \\ \hline
$\bar{c}_z$            &  $W^-$    &     Z    &    $c^+$   &  $g^2\cos^2\theta_W g^{\beta\delta}$     \\ \hline
$\bar{c}^-$        &  $W^+$    &     Z    &     $c_z$    &  $-g^2\cos^2\theta_W g^{\beta\delta}$     
                                                                              \\ \hline
$\bar{c}^+$        &  $W^-$    &     Z    &       $c_z$    &  $g^2\cos^2\theta_W g^{\beta\delta}$     \\ \hline
$\bar{c}^+$        &   Z       &     Z    &    $c^-$   &  $2g^2\cos^2\theta_W g^{\beta\delta}$   \\ \hline
$\bar{c}^+$        &   Z       &     Z    &    $c^+$   &  $-2g^2\cos^2\theta_W g^{\beta\delta}$   \\ \hline
$\bar{c}_z$        &  $W^+$    &     $\gamma$   &    $c^-$   &  $-eg\cos\theta_W g^{\beta\delta}$     \\ \hline
$\bar{c}_z$        &  $W^-$    &     $\gamma$    &    $c^+$   &  $eg\cos\theta_W g^{\beta\delta}$     \\ \hline
$\bar{c}^-$        &  $W^+$    &     $\gamma$    &      $c_z$     &  $-eg\cos\theta_W g^{\beta\delta}$     \\ \hline
$\bar{c}^+$        &  $W^-$    &     $\gamma$    &      $c_z$     &  $eg\cos\theta_W g^{\beta\delta}$     \\ \hline
$\bar{c}_{\gamma}$   &  $W^+$    &     Z    &    $c^-$   &  $-eg\cos\theta_W g^{\beta\delta}$     \\ \hline
$\bar{c}_{\gamma}$&  $W^-$    &     Z    &    $c^+$   &  $eg\cos\theta_W g^{\beta\delta}$     \\ \hline
$\bar{c}^-$        &  $W^+$    &     Z    &      $\gamma$     &  $-eg\cos\theta_W g^{\beta\delta}$     \\ \hline
$\bar{c}^+$        &  $W^-$    &     Z    &      $\gamma$     &  $eg\cos\theta_W g^{\beta\delta}$     \\ \hline
$\bar{c}^+$        &   $\gamma$       &     Z    &    $c^-$   &  $-2 eg\cos\theta_W g^{\beta\delta}$   \\ \hline
$\bar{c}^-$        &   $\gamma$       &     Z    &    $c^+$   &  $2 eg\cos\theta_W g^{\beta\delta}$   \\ \hline
\end{tabular}
\end{center}\begin{center}{Table II}
\end{center}

\subsection{An example of $W^+W^-\to \gamma\gamma$}

The computational simplification by using the GN vertices
results mainly from two sources. First and foremost, each $3g$ vertex
gets reduced from six terms to three terms. Secondly, the vanishing
vertex appearing in the first line of Table II means that
charged-scalar exchange will never occur in tree diagrams
in spite of Feynman propagators being used. The combination of
these two and other minor effects results in enormous
simplification of the algebra.

To illustrate the simplification, 
we have computed the tree diagrams for $W^++W^-\to \gamma+\gamma$
in different gauges.
 The total number of terms in the unitary, Feynman,
 and GN gauges are respectively  147, 77, and 21. 
Using $Mathematica^R$ to compute, the total time needed for each of 
these three cases turns out to be 8.69, 2.13, and 0.52 seconds. 
\begin{center}
\begin{tabular}{|l|l|l|}   \hline
Gauge    & Time  & Number of terms  \\ \hline
Feynman  & 2.13s &     77           \\ \hline
unitary  & 8.69s &     147          \\ \hline
GN       & 0.52s &     21           \\ \hline
\end{tabular}
\end{center}\begin{center}
{Table III}
\end{center}

\section{Conclusion}
In this paper we have shown how to carry out gauge transformation graphically
in the EW theory. The idea and procedure are fairly similar to those used
previously for QCD \cite{YJ}, 
but the presence of spontaneously symmetry breaking
complicates matter and makes the analysis much longer. Using these graphical
rules, new gauges not accessible to the operator or path-integral approach
can be contemplated. Gauges  can now be designed differently for different
sets of Feynman diagrams to maximize the reduction of the algebra
caused by the gauge-dependent terms. As an illustration of the graphical
rules, we worked out the Gervais-Neveu gauge in the EW theory, and applied
it to calculate the tree process $W^++W^-\to\gamma+\gamma$. The computation
in the GN gauge is a factor of 4 faster than in the Feynman gauge, and
a factor of 17 faster than in the unitary gauge.

\section{Acknowledgement}
This research was supported in part by the Natural Science and 
Engineering Research Council of Canada and by the Qu\'{e}bec Department
of Education. Y.J.F. acknowledges the support of the Carl Reinhardt Major 
Foundation. 

\newpage
\appendix

\section{Feynman rules in Electroweak theory}
\begin{picture}(100,200) (0,0)
\put (10,60){\epsfbox {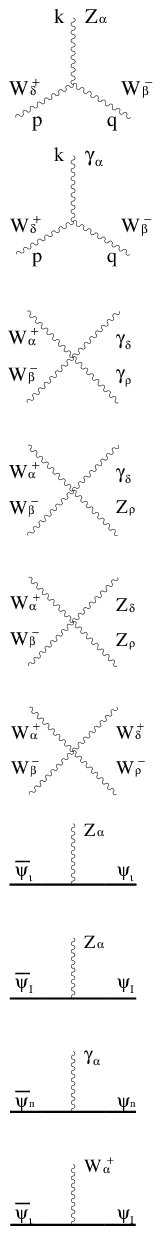}}

\put (80, 190){$[g \cos \theta_W]\left((k-p)^{\delta}g^{\alpha\beta}+(p-q)^{\alpha}g^{\beta\delta}+(q-k)^{\beta}g^{\alpha\delta}\right)$}

\put (80, 145){$[e] \left((k-p)^{\delta}g^{\alpha\beta}+(p-q)^{\alpha}g^{\beta\delta}+(q-k)^{\beta}g^{\alpha\delta}\right)$}

\put (80,100) {$e^2\left(g^{\alpha\rho}g^{\beta\delta}+g^{\alpha\delta}g^{\beta\rho}-2 g^{\alpha\beta}g^{\rho\delta}\right)$}

\put (80,62) {$e g \cos\theta_W\left(g^{\alpha\rho}g^{\beta\delta}+g^{\alpha\delta}g^{\beta\rho}-2 g^{\alpha\beta}g^{\delta\rho}\right)$}

\put (80,24) {$g^2 \cos^2\theta_W\left(g^{\alpha\rho}g^{\beta\delta}+g^{\alpha\delta}g^{\beta\rho}-2 g^{\alpha\beta}g^{\delta\rho}\right)$}

\put (80, -14) {$-g^2 \left(g^{\alpha\rho}g^{\beta\delta}+g^{\alpha\beta}g^{\delta\rho}-2 g^{\alpha\delta}g^{\beta\rho}\right)$}

\put (80,-205) {$[{ig\over  2}](q-p)_{\alpha}$}
\put (80,-237.5) {$[{ig\over  2}](q-p)_{\alpha}$}
\put (80,-270) {$[{g\over  2}] (q-p)_{\alpha}$}
\put (80,-302.5){$[{g\over  2}] (p-q)_{\alpha}$}
\put (80,-335){$[{g\over  2 \cos\theta_W}](1-2 \cos^2\theta_W) (q-p)_{\alpha}$}
\put (80,-367.5){$[e] (p-q)_{\alpha}$}
\put (80,-400){$[{ig\over  \cos\theta_W}](q-p)_{\alpha}$}

\put (80,-172.5){${g\over  2\sqrt{2}}\gamma_{\alpha}U_{iI}(1-\gamma_5)$}
\put (80,-140){${g\over  2\sqrt{2}}\gamma_{\alpha}U_{iI}^*(1-\gamma_5)$}
\put (80,-107.5){$e Q_n\gamma_{\alpha}\ (n=i,I)$}
\put (80,-75){${g\over  2\cos \theta_W}\gamma_{\alpha}({1\over  2}-2 Q_I\sin^2\theta_W-{1\over  2} \gamma_5)$}
\put (80,-42.5){$-{g\over  2\cos \theta_W}\gamma_{\alpha}({1\over  2}-2 Q_i\sin^2\theta_W-{1\over  2} \gamma_5)$}

\put (0,210){\line(50,0){350}}
\put (0,165){\line(50,0){350}}
\put (0,123){\line(50,0){350}}
\put (0,81){\line(50,0){350}}
\put (0,43){\line(50,0){350}}
\put (0,5){\line(50,0){350}}
\put (0,-30){\line(50,0){350}}
\put (0,-60){\line(50,0){350}}
\put (0,-90){\line(50,0){350}}
\put (0,-122.5){\line(50,0){350}}
\put (0,-155){\line(50,0){350}}
\put (0,-187.5){\line(50,0){350}}
\put (0,-223){\line(50,0){350}}
\put (0,-255.5){\line(50,0){350}}
\put (0,-288){\line(50,0){350}}
\put (0,-320.5){\line(50,0){350}}
\put (0,-353){\line(50,0){350}}
\put (0,-385){\line(50,0){350}}
\put (0,-418){\line(50,0){350}}

\put (0,-418){\line(0,100){628}}
\put (350,-418){\line(0,100){628}}
\end{picture}

\newpage
\begin{picture}(100,200) (0,0)
\put (10,60){\epsfbox {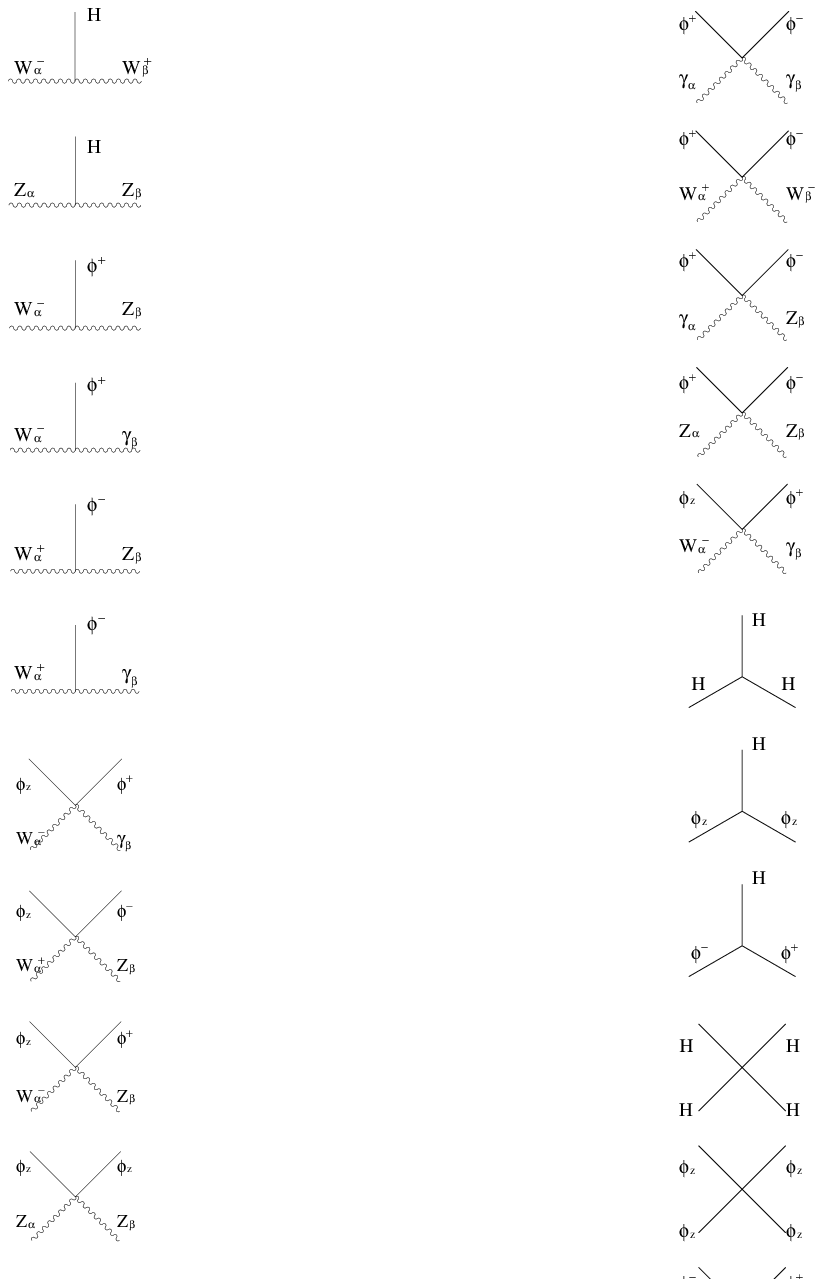}}

\put (80,190) {$g M_Wg^{\alpha\beta}$}
\put (80,154.7) {${g M_z\over  \cos\theta_W}g^{\alpha\beta}$}
\put (80,119.4) {$ie M_z\sin\theta_Wg^{\alpha\beta}$}
\put (80,84) {$-ie M_Wg^{\alpha\beta}$}
\put (80,49) {$-ieM_z\sin\theta_W g^{\alpha\beta}$}
\put (80,13.5) {$i e M_W g^{\alpha\beta}$}

\put (80,-23.5) {$-{eg\over  2}g^{\alpha\beta}$}
\put (80,-61) {${e^2\over  \cos\theta_W}g^{\alpha\beta}$}
\put (80,-98) {${e^2\over  \cos\theta_W}g^{\alpha\beta}$}
\put (80,-135) {${g^2\over  \cos^2\theta_W}g^{\alpha\beta}$}
\put (80,-172) {${g^2\over  2}g^{\alpha\beta}$}
\put (80,-209) {${ig^2\over  2}g^{\alpha\beta}$}
\put (80,-246) {$-{ig^2\over  2}g^{\alpha\beta}$}
\put (80,-283) {$-{i e^2\over  2\cos\theta_W} g^{\alpha\beta}$}
\put (80,-319.5) {${i e^2\over  2\cos\theta_W} g^{\alpha\beta}$}
\put (80,-356.5) {${g^2\over  2\cos^2\theta_W}g^{\alpha\beta}$}
\put (80,-393.5) {${g^2\over  2}g^{\alpha\beta}$}

\put (270,190) {$2 e^2 g^{\alpha\beta}$}
\put (270,154.7) {${g^2\over  2} g^{\alpha\beta}$}
\put (270,119.4) {${eg(2 \sin^2\theta_W-1)\over \cos\theta_W}g^{\alpha\beta}$}
\put (270,84) {${(2\sin^2\theta_W-1)g^2\over  2\cos^2\theta_W}g^{\alpha\beta}$}
\put (270,49) {$-{eg\over  2}g^{\alpha\beta}$}

\put (270,13.5) {$-{3g M_H^2\over  2 M_z \cos\theta_W}$}
\put (270,-23.5) {$-{g M_H^2\over  2 M_z \cos\theta_W}$}
\put (270,-61) {$-{g M_H^2\over  2 M_z\cos\theta_W}g^{\alpha\beta}$}

\put (270,-98) {$-{3 g^2 M_H^2\over  4M_z^2\cos^2\theta_W}$}
\put (270,-135) {$-{3 g^2 M_H^2\over  4M_z^2\cos^2\theta_W}$}
\put (270,-172) {$-{ g^2 M_H^2\over  4M_z^2\cos^2\theta_W}$}
\put (270,-209) {$-{ g^2 M_H^2\over  4M_z^2\cos^2\theta_W}$}

\put (270,-246) {$-{g^2 M_H^2\over  4M_z^2\cos^2\theta_W}$}
\put (270,-283) {$-{ g^2 M_H^2\over  2M_z^2\cos^2\theta_W}$}

\put (250,-319.5) {${i g\over 2}M_z(2\cos^2\theta_W-1)$}
\put (270,-356.5) {$ieM_W$}
\put (270,-393.5) {$-i {g\over 2} M_z$}

\put (0,210){\line(50,0){350}}
\put (0,177.5){\line(50,0){350}}
\put (0,142.2){\line(50,0){350}}
\put (0,107){\line(50,0){350}}
\put (0,71.5){\line(50,0){350}}
\put (0,36.5){\line(50,0){350}}
\put (0,1.5){\line(50,0){350}}
\put (0,-42){\line(50,0){350}}
\put (0,-79){\line(50,0){350}}
\put (0,-116){\line(50,0){350}}
\put (0,-153){\line(50,0){350}}
\put (0,-190){\line(50,0){350}}
\put (0,-227){\line(50,0){350}}
\put (0,-264){\line(50,0){350}}
\put (0,-301){\line(50,0){350}}
\put (0,-338){\line(50,0){350}}
\put (0,-375){\line(50,0){350}}
\put (0,-412){\line(50,0){350}}

\put (0,-412){\line(0,100){622}}
\put (350,-412){\line(0,100){622}}
\put (175,-412){\line(0,100){622}}
\end{picture}
\newpage

\begin{picture}(100,200) (0,0)
\put (10,60){\epsfbox {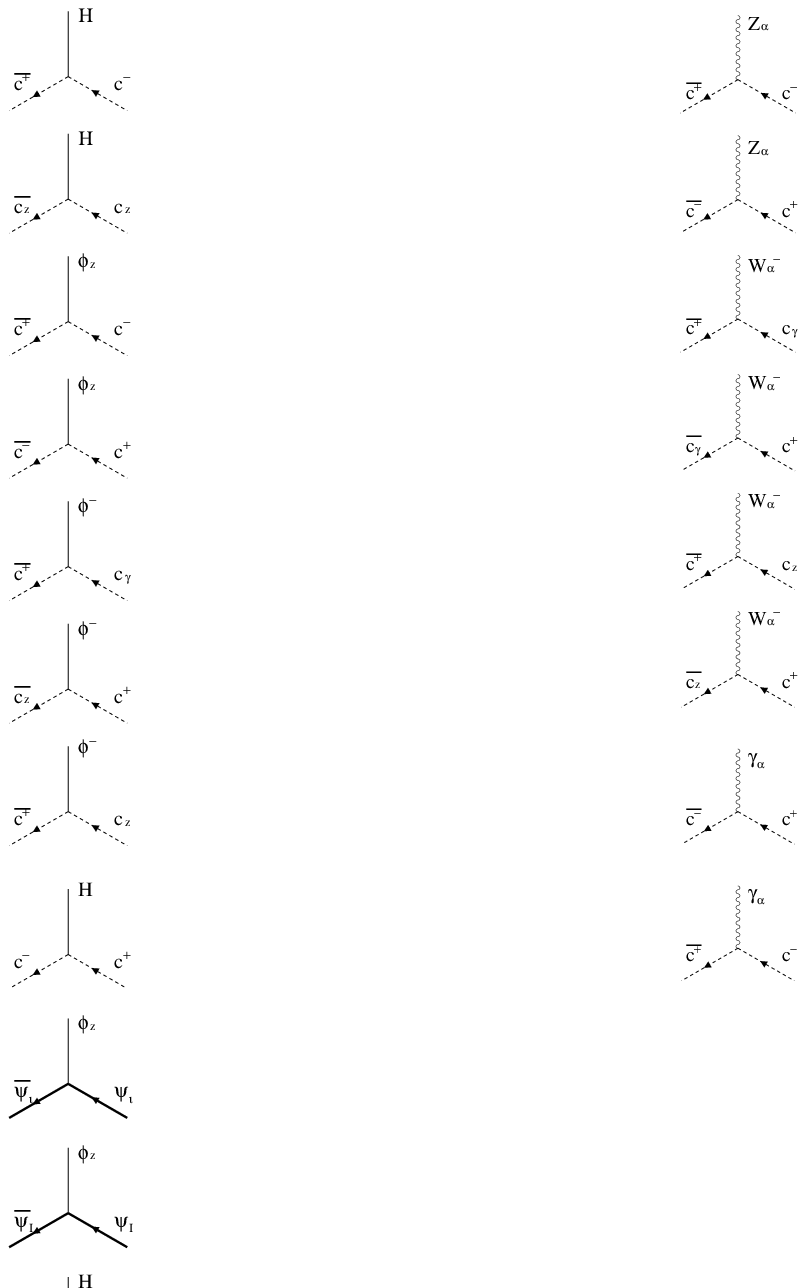}}

\put (80,190) {$-{g\over 2} M_W$}
\put (80,154.7) {$-{g\over 2\cos\theta_W}M_z$}
\put (80,119.4) {${ig\over 2}M_W$}
\put (80,84) {$-{ig\over 2} M_W$}
\put (80,49) {$-ie M_W$}
\put (80,13.5) {$i{g\over 2}M_z $}
\put (75,-23.5) {$-i{g\over 2} M_z(2\cos^2\theta_W-1)$}
\put (80,-61) {$-{g\over  2} M_W$}

\put (80,-98) {$i{g\over 2M_W} m_i$}
\put (80,-135) {$i{g\over 2M_W} m_I$}
\put (80,-172) {$-i {g\over 2M_W}m_n \ (n=i,\ I)$}
\put (80,-209) {$-i{g\over 2\sqrt{2}M_W}U_{iI}((m_i-m_I)+(m_i+m_I)\gamma_5)$}
\put (80,-246) {$-i{g\over 2\sqrt{2}M_W}U_{iI}^*((m_I-m_i)+(m_i+m_I)\gamma_5)$}

\put (80,-283) {$[e] p_{\alpha}$}
\put (80,-319.5) {$-[e] p_{\alpha}$}
\put (80,-356.5) {$[g\cos\theta_W] p_{\alpha}$}
\put (80,-393.5) {$-[g\cos\theta_W] p_{\alpha}$}

\put (270,190) {$[g\cos\theta_W] p_{\alpha}$}
\put (270,154.7) {$-[g\cos\theta_W] p_{\alpha}$}
\put (270,119.4) {$-[e] p_{\alpha}$}
\put (270,84) {$[e] p_{\alpha}$}
\put (270,49) {$-[g\cos\theta_W] p_{\alpha}$}
\put (270,13.5) {$[g\cos\theta_W] p_{\alpha}$}
\put (270,-23.5) {$-[e] p_{\alpha}$}
\put (270,-61) {$[e] p_{\alpha}$}

\put (0,210){\line(50,0){350}}
\put (0,177.5){\line(50,0){350}}
\put (0,142.2){\line(50,0){350}}
\put (0,107){\line(50,0){350}}
\put (0,71.5){\line(50,0){350}}
\put (0,36.5){\line(50,0){350}}
\put (0,1.5){\line(50,0){350}}
\put (0,-42){\line(50,0){350}}
\put (0,-79){\line(50,0){350}}
\put (0,-116){\line(50,0){350}}
\put (0,-153){\line(50,0){350}}
\put (0,-190){\line(50,0){350}}
\put (0,-227){\line(50,0){350}}
\put (0,-264){\line(50,0){350}}
\put (0,-301){\line(50,0){350}}
\put (0,-338){\line(50,0){350}}
\put (0,-375){\line(50,0){350}}
\put (0,-412){\line(50,0){350}}

\put (0,-412){\line(0,100){622}}
\put (350,-412){\line(0,100){622}}
\put (175,-79){\line(0,100){289}}
\end{picture}
\newpage
\section{Divergence relations and cancellations}
This appendix deals with all the graphical identities in the Feynman gauge.
Those involving fermions can be obtained similarly but will not be 
explicitly discussed here.
Wavy, dotted, solid lines are gauge, ghost, and scalar
fields respectively, and a cross represents a divergence. Moreover,

\begin{enumerate}
\item When a ghost line is connected to its unphysical scalar, 
a factor $-iM$ is assigned to the two-point vertex, where $M$
is the mass of the corresponding gauge particle.

\item The sign and the numerical factor for a sliding diagram
is determined as follows. The numerical factor is given by the
`coupling constant' of the three-point vertex where the sliding
diagram comes from. By `coupling constant', one means the numerical
factor at the vertex in App.~A enclosed in a square bracket $[\cdots]$.
The sign is determined as follows. First, for vertices not involving a
ghost, a standard clockwise orientation
is established for each of them. They are,
in clockwise orders, 
$W^-W^+\gamma(Z)$, $\phi^{\pm}W^{\mp}H$, 
$\phi^{+}W^{-}\phi_z$, $\phi_zW^+\phi^-$, $\phi^+Z\phi^-$, 
$\phi^-\gamma\phi^+$, and $\phi_zZH$. If the wandering ghost line turns
to the left along the direction of its arrow, then
the sign is negative; if it turns  to the right, then the sign
is positive. For vertices involving ghost lines, there is only one
sliding diagram per vertex, then the sign is positive for the vertices
$\bar{G}_W W^-G_z$, $\bar{G}_W W^-G_\gamma$, $\bar{G}_z W^+G_W$, 
$\bar{G}_{\gamma}W^+G_W$, $\bar{G}_W\gamma G_{W^+}$, $\bar{G}_W ZG_{W^+}$, and
negative for the vertices
$\bar{G}_WW^+G_z$, $\bar{G}_WW^+G_\gamma$, $\bar{G}_zW^-G_W$, 
$\bar{G}_{\gamma} W^-G_W$, $\bar{G}_W\gamma G_{W^-}$, $\bar{G}_WZG_{W^-}$.
\end{enumerate}

\subsection{Divergence relations}
\begin{enumerate}
\item ggg:

The relation is shown in Fig.~8.

\begin{figure}
\vskip -0 cm
\centerline{\epsfxsize 4.7 truein \epsfbox {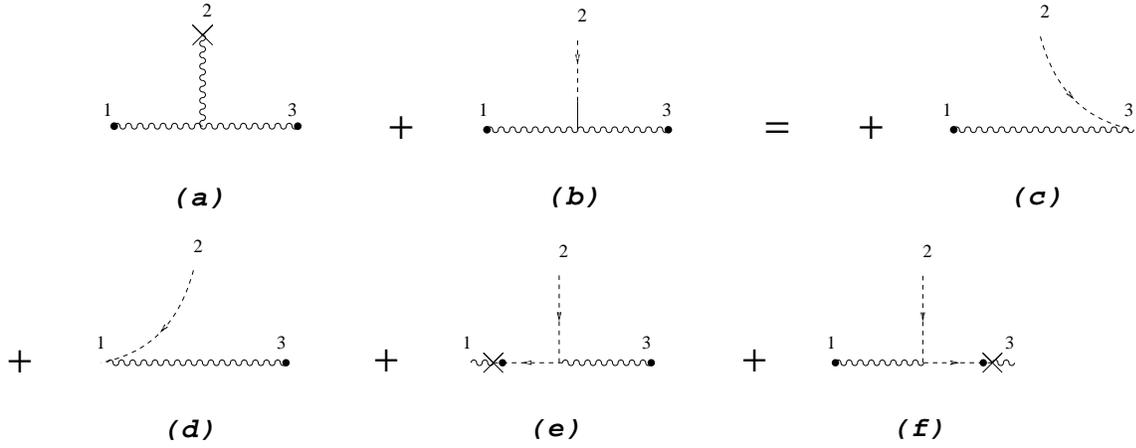}}
\nobreak
\vskip -11cm\nobreak
\vskip .1cm
\caption{Divergence relation of the triple gauge boson vertex. If line 2 is
a photon line or  $Z$, then (b) is absent.}
\end{figure}

\item ggS:

There are several cases.

\begin{enumerate}
\item $W\gamma\phi$ with the cross on the $\gamma$ line, 
and $WZ\phi$ with the cross on the $Z$ line are shown in Fig.~9.

\begin{figure}
\vskip -0 cm
\centerline{\epsfxsize 4.7 truein \epsfbox {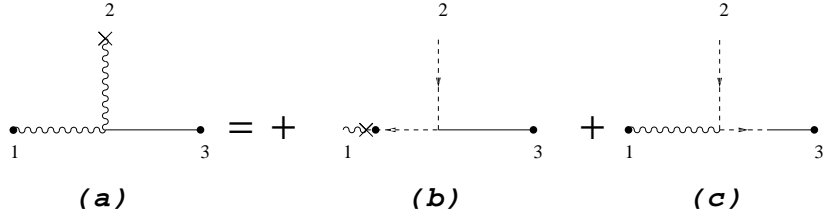}}
\nobreak
\vskip -13.5 cm\nobreak
\vskip .1 cm
\caption{Divergence relation of $ggS$.}
\end{figure}

\item $W\gamma\phi$ with the cross on the W line. It is shown in Fig.~10 with 
10(d) absent.

\begin{figure}
\vskip -4 cm
\centerline{\epsfxsize 4.7 truein \epsfbox {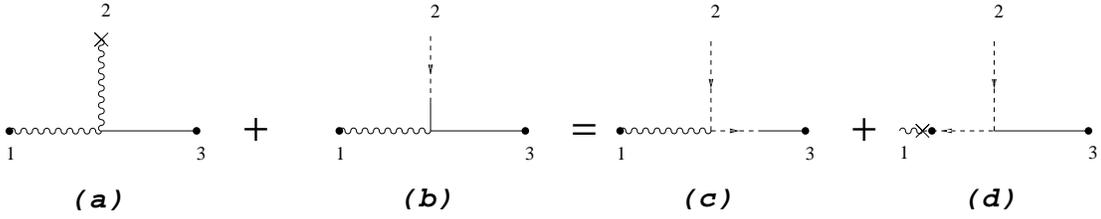}}
\nobreak
\vskip -10 cm\nobreak
\vskip .1 cm
\caption{The crossed line in (a) is a W particle.}
\end{figure}

\item $WWH$ and $ZZH$ vertices.
The situations here are the same as in Fig.~10 with 10(c) absent.

\end{enumerate}

\item gSS:
Because of the presence of 
the problem of mass compensation, we have to discuss them separately.
\begin{enumerate}
\item $Z\phi\phi$ and $\gamma\phi\phi$.
All of them are similar to the scalar QED case, and they are shown in Fig.~11.

\begin{figure}
\vskip -1 cm
\centerline{\epsfxsize 4.7 truein \epsfbox {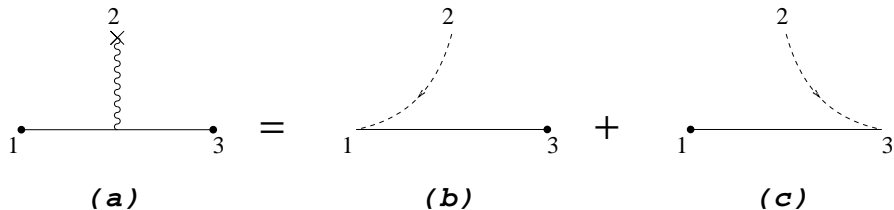}}
\nobreak
\vskip -13.5 cm\nobreak
\vskip .1 cm
\caption{The divergence relation of $gSS$ vertex.}
\end{figure}

\item $g\phi H$:
The W line can be either an outgoing $W^-$ or an outgoing $W^+$. 
We use Fig.~12 to cover these two possibilities by choosing 
the following convention. If the cross is on an outgoing $W^+$, 
we use line1 to represent the Higgs. If the cross is on an outgoing
$W^-$, we choose line3 to be the Higgs.

\begin{figure}
\vskip -0 cm
\centerline{\epsfxsize 4.7 truein \epsfbox {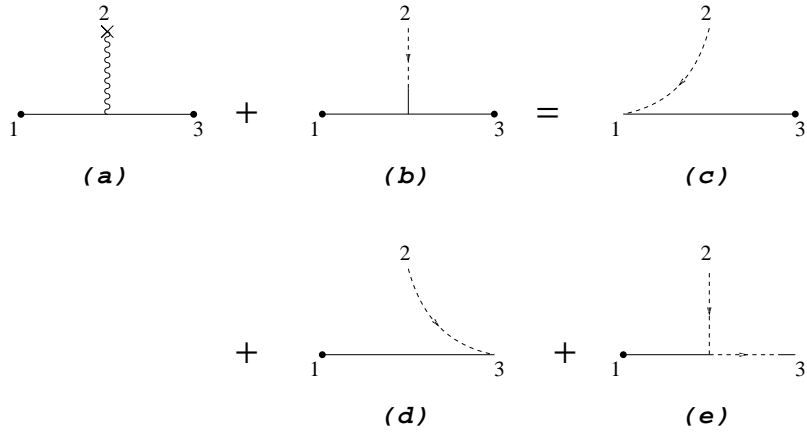}}
\nobreak
\vskip -11.5 cm\nobreak
\vskip .1 cm
\caption{Another divergence relation of the $gSS$ vertex.}
\end{figure}

\item $g\phi \phi_z$:
The ghost now can turn to either of the scalar lines as shown in Fig.~13.
\begin{figure}
\vskip -3.5 cm
\centerline{\epsfxsize 4.7 truein \epsfbox {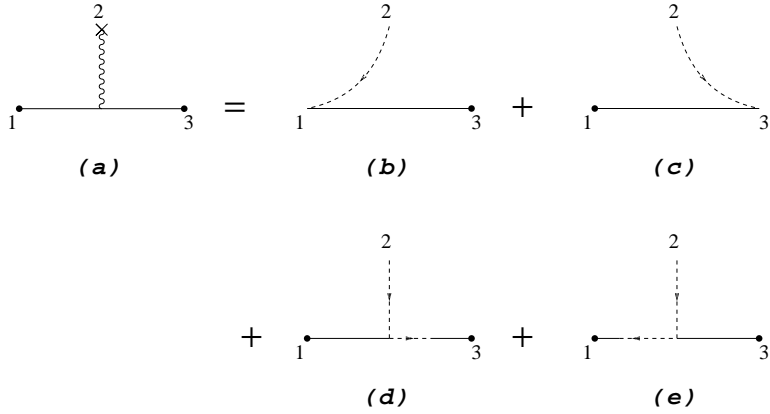}}
\nobreak
\vskip -8 cm\nobreak
\vskip .1 cm
\caption{Divergence relation of $gSS$ vertex again.}
\end{figure}
\end{enumerate}

\item gggg: 

All the cases can be represented by one identity as shown in 
Fig.~14.
\begin{figure}
\vskip -3.5 cm
\centerline{\epsfxsize 4.7 truein \epsfbox {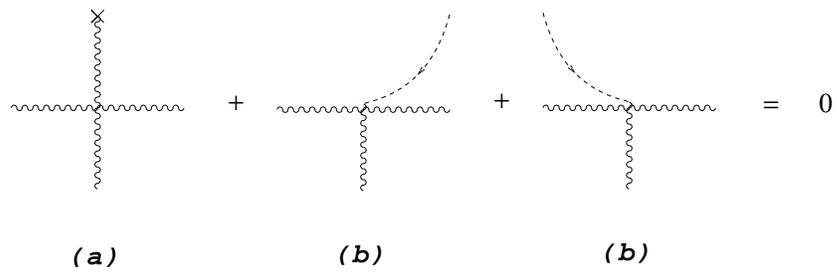}}
\nobreak
\vskip -17.5 cm\nobreak
\vskip .1 cm
\caption{ The divergence relation for $gggg$ vertex.}
\end{figure}

\item ggSS:

There are three possibilities here: the boson can slide along gS, SS, gSS.
\begin{enumerate}
\item gS:
They include $W\gamma H\phi$, $W\gamma\phi\phi_z$, $WW\phi\phi$, 
$ZZ\phi\phi$, $\gamma\gamma\phi\phi$,  and $\gamma Z\phi\phi$. 
This is shown in Fig.~15.
\begin{figure}
\vskip -0.5 cm
\centerline{\epsfxsize 4.7 truein \epsfbox {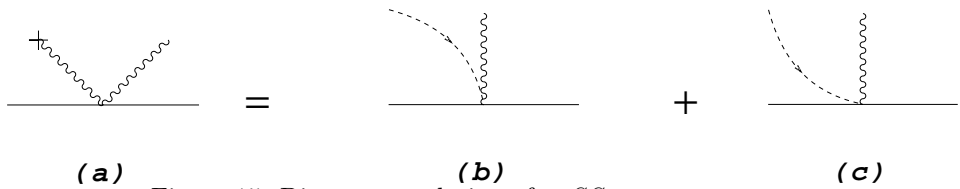}}
\nobreak
\vskip -13 cm\nobreak
\vskip .1 cm
\caption{Divergence relation of $ggSS$ vertex.}
\end{figure}

\item SS:
They include $WWHH$, $ZZHH$, $WW\phi_z\phi_z$, and $ZZ\phi_z\phi_z$. We use 
Fig.~16 to show them.
\begin{figure}
\vskip -0 cm
\centerline{\epsfxsize 4.7 truein \epsfbox {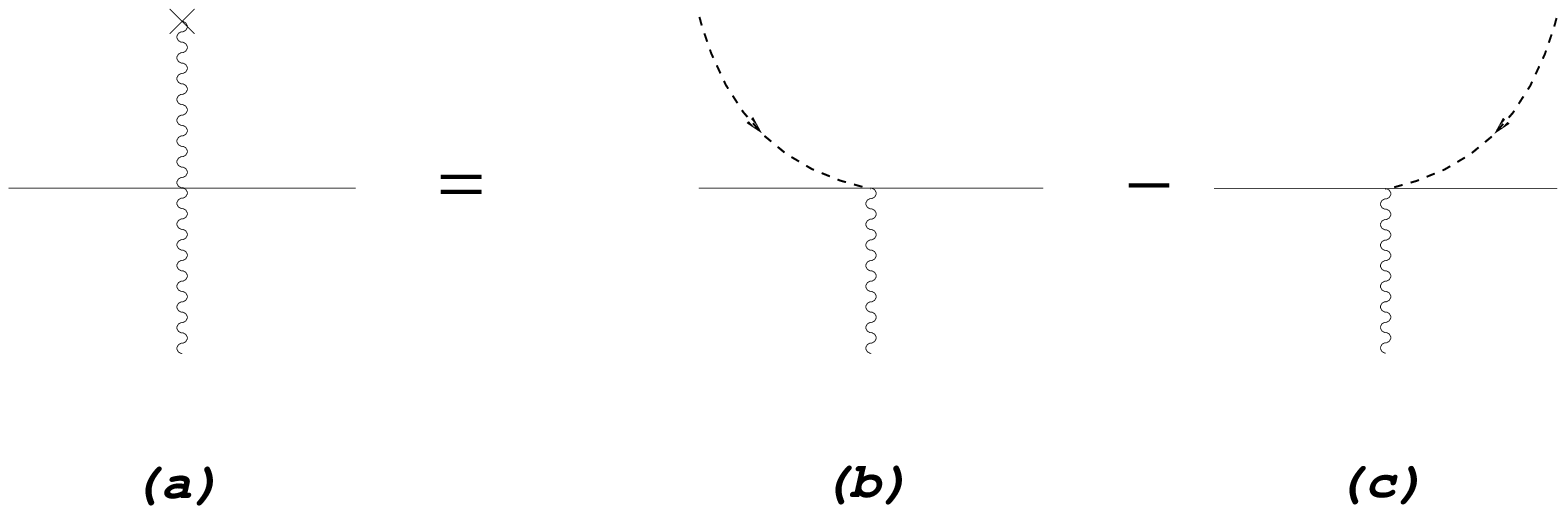}}
\nobreak
\vskip -13 cm\nobreak
\vskip .1 cm
\caption{This figure covers two cases: (1) line 2 and line 4 in 
(a) are all scalar, while line 1 and line 3 are Higgs. 
(2) All of them are scalar while line 1 and line 3 have the   same charge. }
\end{figure}

\item gSS:
They include $WZH\phi$, and $WZ\phi\phi_z$. We use Fig.~17 to show them.
\begin{figure}
\vskip -1.5 cm
\centerline{\epsfxsize 4.7 truein \epsfbox {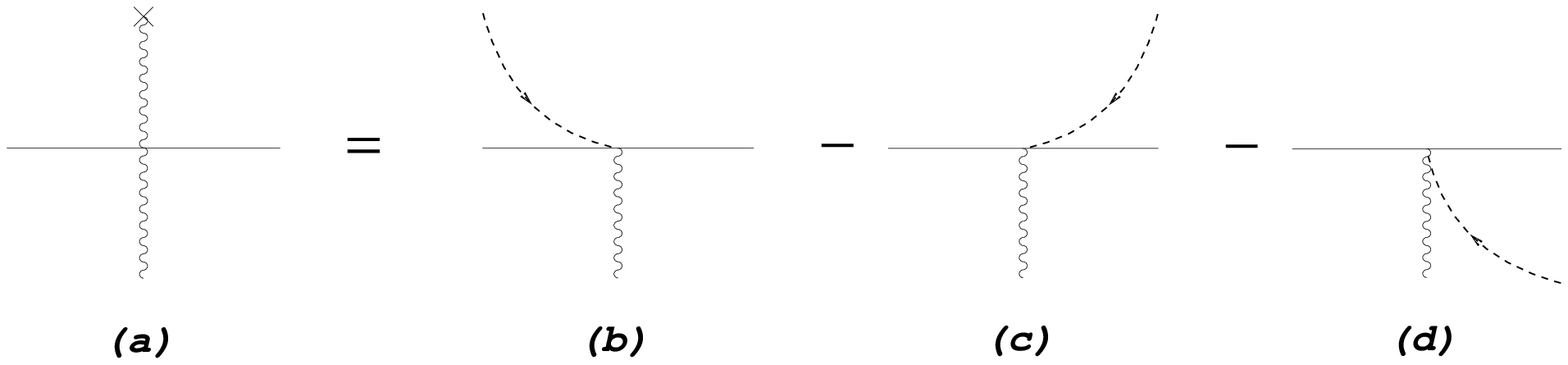}}
\nobreak
\vskip -12 cm\nobreak
\vskip .1 cm
\caption{Divergence relation of $ggSS$ vertex. }
\end{figure}

\end{enumerate}

\item gGG:

The number of compensation diagrams on the left varies:
\begin{enumerate}
\item
For vertices
$W{\bar G_\phi} G_z$, $Z{\bar G_\phi} G_\phi$, and $W{\bar G_z}G_\phi$,
there are two compensation diagrams as shown in Fig.~18.

\begin{figure}
\vskip -0 cm
\centerline{\epsfxsize 4.7 truein \epsfbox {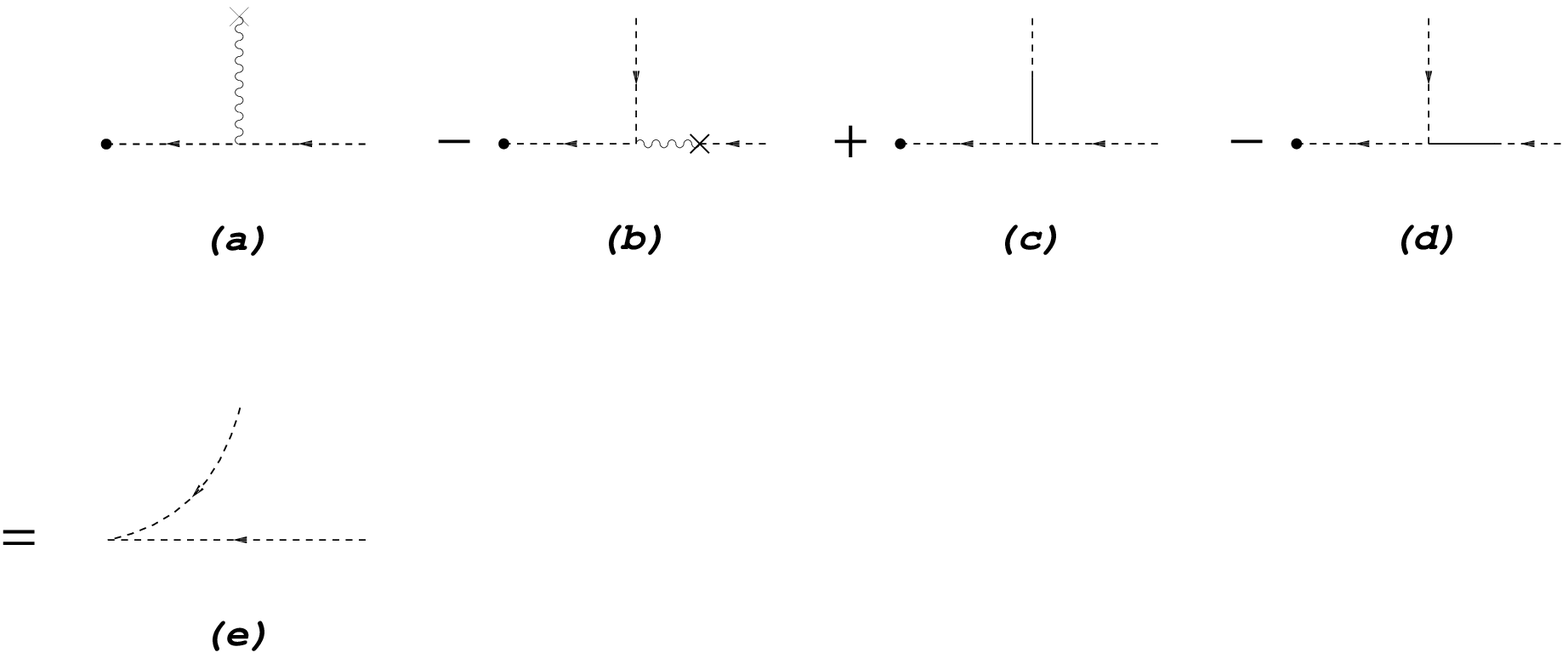}}
\nobreak
\vskip -10.5 cm\nobreak
\vskip .1 cm
\caption{ Divergence relation of $gGG$ vertex.}
\end{figure}

\item For vertices
$W{\bar G_\phi} G_\gamma$, and $\gamma{\bar G_\phi} G_\phi$, 
there is one compensation diagram as shown in Fig.~19.

\begin{figure}
\vskip -1 cm
\centerline{\epsfxsize 4.7 truein \epsfbox {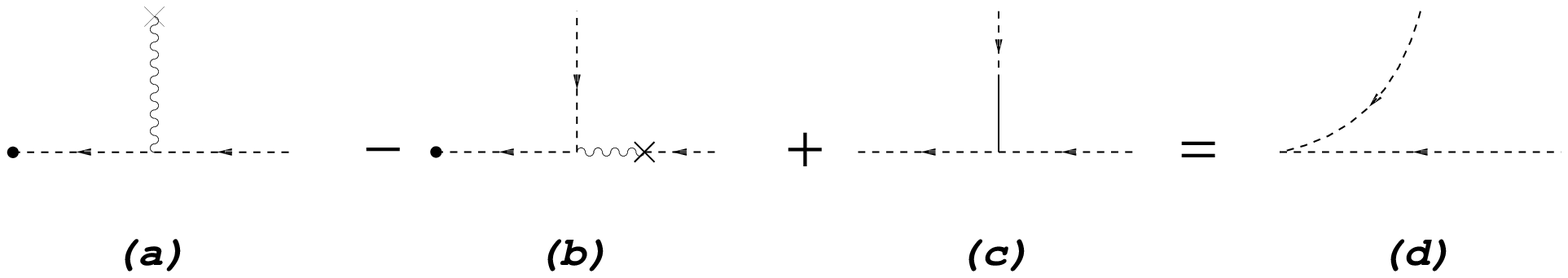}}
\nobreak
\vskip -12.5 cm\nobreak
\vskip .1 cm
\caption{ Divergence relation of the $gGG$ vertex.}
\end{figure}
\item For the vertex
$W{\bar G_\gamma}G_\phi$, there is no compensation diagram as shown in Fig.~20.
\end{enumerate}

\begin{figure}
\vskip -0 cm
\centerline{\epsfxsize 4.7 truein \epsfbox {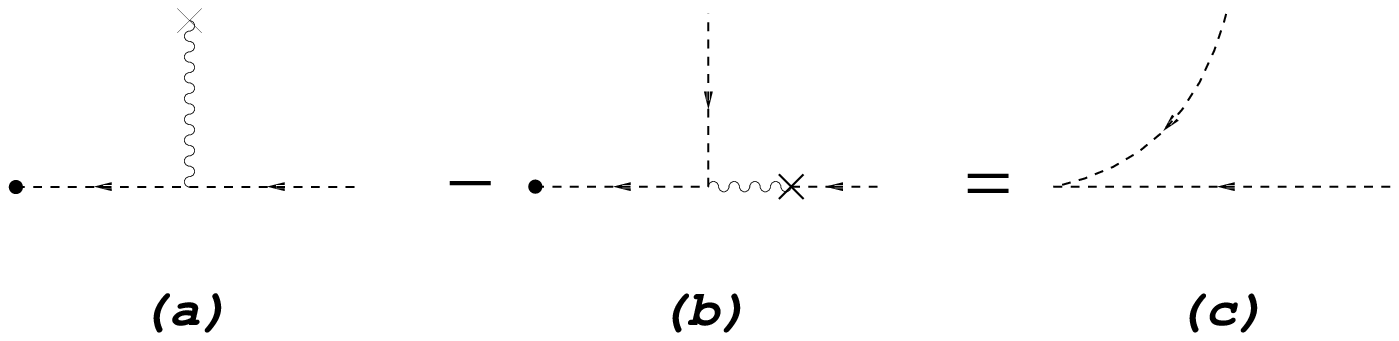}}
\nobreak
\vskip -13.5 cm\nobreak
\vskip .1 cm
\caption{ Divergence relation of the $gGG$ vertex.}
\end{figure}
\end{enumerate}

These are all the divergence relations in Feynman gauge.

\subsection{Cancellation of the sliding diagrams}
\begin{enumerate} 
\item ggg vertices:  They are the same as in QCD diagrams.
\item ggS vertices:
The various cases are summarized in the following table and Figs.~21 to 24.

\begin{center}
\begin{tabular}{|l|l|l|l|l|l|l|}   \hline
Vertex       & sliding line  & figure    &   1   &   2     &   3  &  4   \\   \hline
$HWW$        & $\gamma,\ Z$        &  Fig.~21    &  $W$  & $\gamma,\ Z$   & $W$  &  $H$ \\   \hline
$HZZ$, $\phi WZ$ & $W$  & Fig.~22   & $Z$   &   $H$  &   $W$    &  $W$  \\ \hline
$W^{\pm}Z(\gamma)\phi^{\mp}$   & $\gamma$      & Fig.~23   & $W$   &   $Z$  &   $\phi$ &  $\gamma$   \\ \cline{2-7}
             & $Z$      & Fig.~24   & $W$   &   $Z$  &   $\phi$ &  $Z$   \\ \cline{2-7}
             & $W^{\mp}$    & Fig.~23   & $W^{\pm}$   &   $W^{\pm}$  &   $\phi^{\mp}$ &  $W^{\mp}$   \\ \hline 
\end{tabular}
\end{center}

\begin{figure}
\vskip 0 cm
\centerline{\epsfxsize 4.7 truein \epsfbox {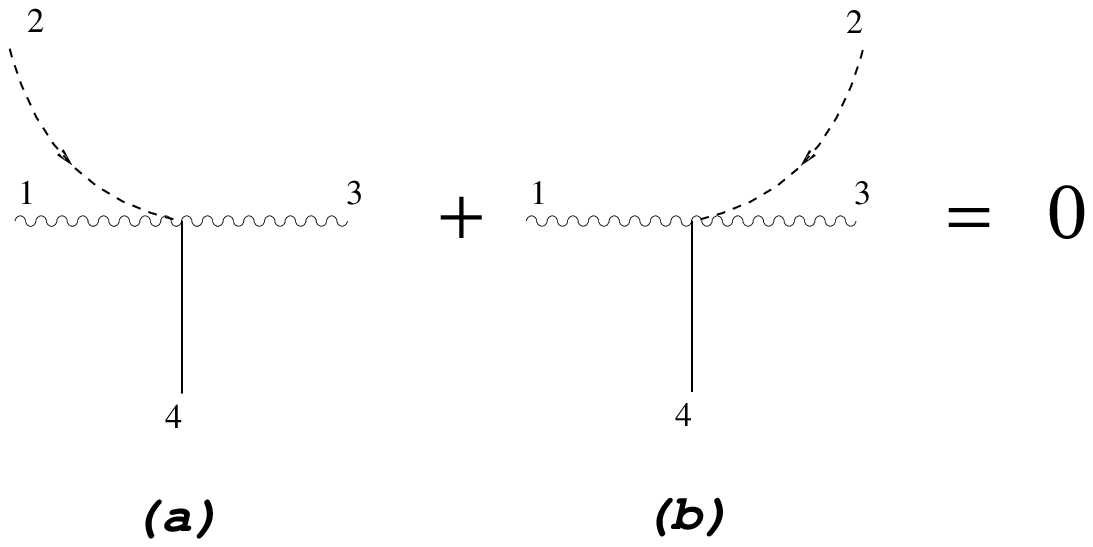}}
\nobreak
\vskip -12.5 cm\nobreak
\vskip .1 cm
\caption{Cancellation relation of the $ggS$ vertex. }
\end{figure}

\begin{figure}
\vskip -1 cm
\centerline{\epsfxsize 4.7 truein \epsfbox {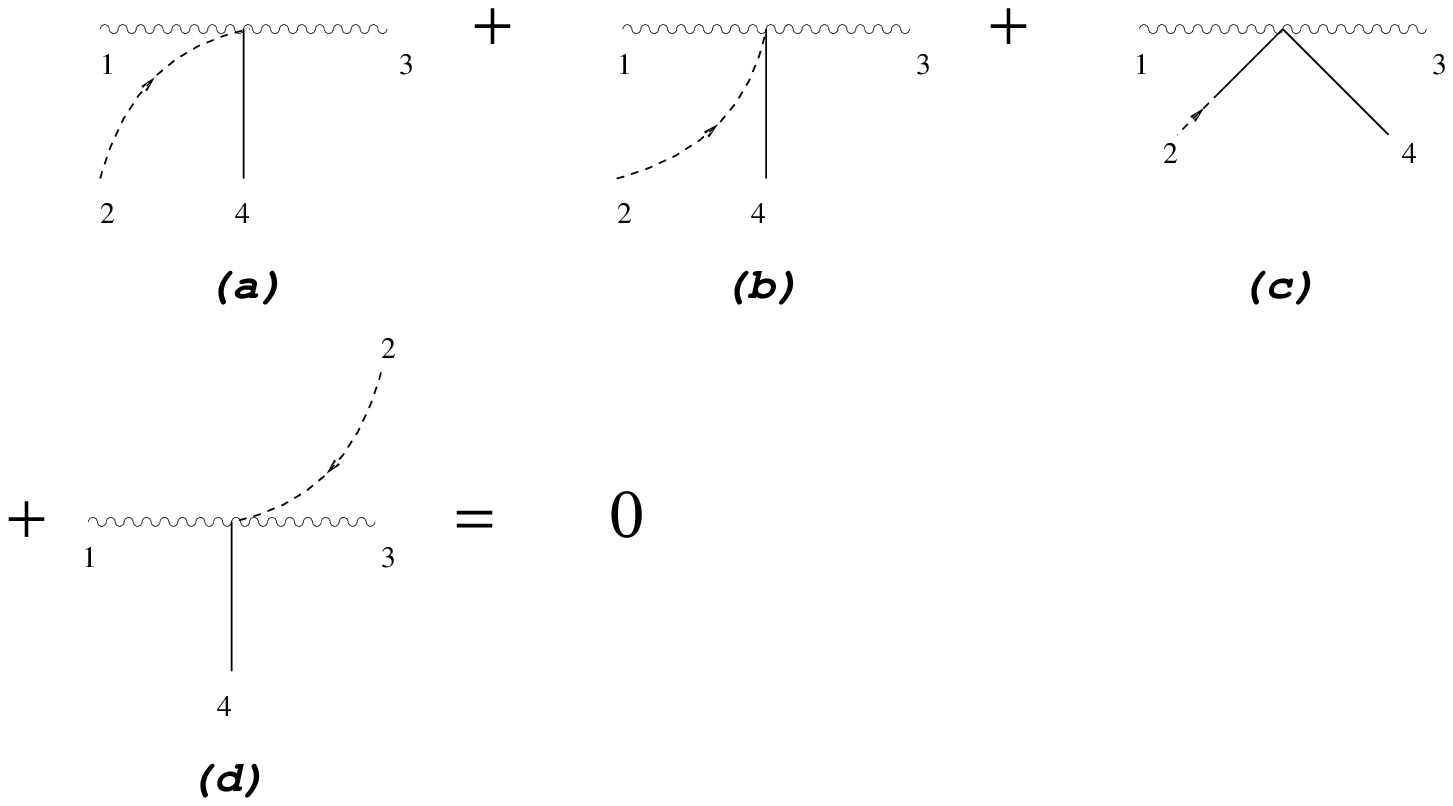}}
\nobreak
\vskip -10.5 cm\nobreak
\vskip .1 cm
\caption{ Cancellation relation of the $ggS$ vertex.}
\end{figure}

\begin{figure}
\vskip -1.2 cm
\centerline{\epsfxsize 4.7 truein \epsfbox {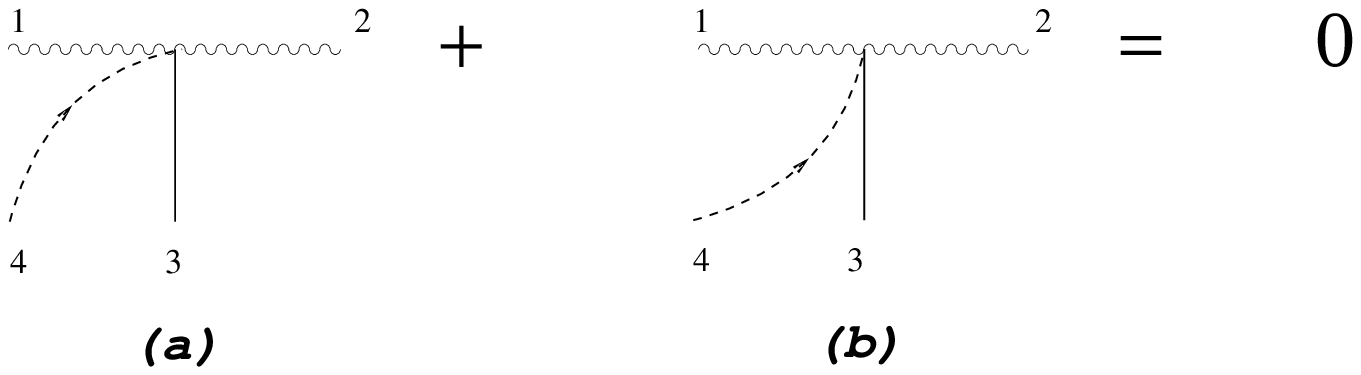}}
\nobreak
\vskip -13 cm\nobreak
\vskip .1 cm
\caption{ Cancellation relation of the $ggS$ vertex.}
\end{figure}

\begin{figure}
\vskip -1 cm
\centerline{\epsfxsize 4.7 truein \epsfbox {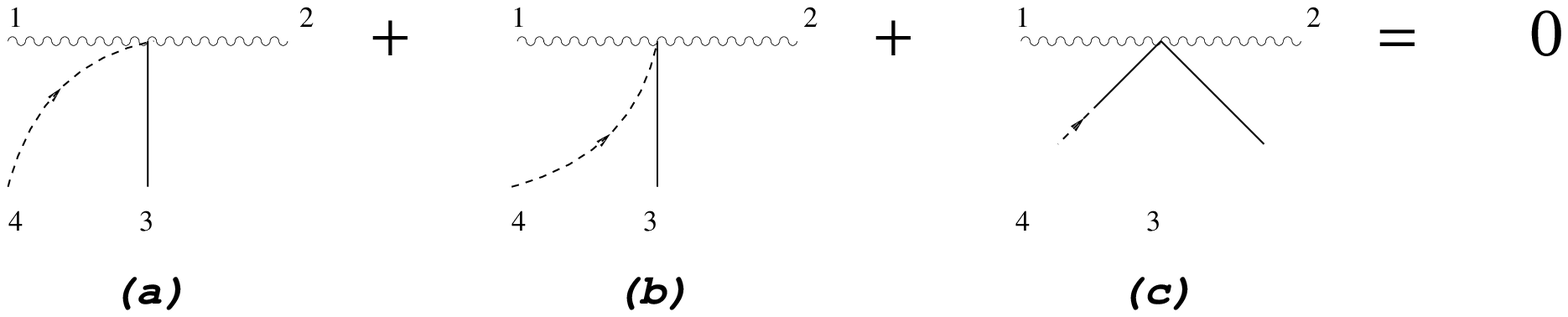}}
\nobreak
\vskip -12.5 cm\nobreak
\vskip .1 cm
\caption{ Cancellation relation of the $ggS$ vertex.}
\end{figure}

\item ggSS vertices:
The various cases are summarized in the following table and Figs.~25 to 30.
\begin{center}
\begin{tabular}{|l|l|l|l|l|l|l|}   \hline
vertex  & sliding line  & figure  &    1   &    2   &    3   &   4    \\ \hline
$HHWW$  & $\gamma,Z$         & Fig.~25   &   $H$  &   $W$  &   $W$  &  $H$   \\ \cline{2-7}
        & $Z$           & Fig.~27   &   $H$  &   $W$  &   $W$  &  $\phi$\\ \cline{2-7}
        & $W$           & Fig.~29   &   $H$  &   $W$  &   $W$  &  $\phi$\\ \hline
$HHZZ$  & $W$           & Fig.~30   &   $H$  &   $Z$  &   $W$  &  $H$   \\ \hline
$H\phi WZ$& $\gamma$         & Fig.~26   &   $H$  &   $Z$  &   $W$  &  $\phi$\\ \cline{2-7}
        &  $Z$          & Fig.~29   &   $H$  &   $W$  &   $Z$  &  $\phi$ \\ \cline{2-7}
        &  $W$          & Fig.~29   &   $H$  &   $W$  &   $W$  &  $\phi$ \\ \cline{2-7}
        &  $W^{\pm}$        & Fig.~25   &   $H$  &   $W^{\mp}$  &   $W^{\mp}$  &  $\phi^{\pm}$ \\ \hline
$HW\gamma\phi$& $\gamma$          & Fig.~26   &   $H$  &   $\gamma$  &   $W$  &  $\phi$ \\ \cline{2-7}
        &  $Z$          & Fig.~29   &   $H$  &   $W$  &   $\gamma$  &  $\phi$ \\ \hline
$WW\phi_z\phi_z$&$\gamma,\ Z$& Fig.~25   &$\phi_z$&   $W$  &   $W$  &$\phi_z$ \\ \cline{2-7}
        &  $W^{\pm}$        & Fig.~29   &  $\phi$&   $W^{\mp}$  &  $W^{\pm}$ &$\phi_z$ \\ \hline
$ZZ\phi_z\phi_z$&$W^{\pm}$  & Fig.~30   &$\phi_z$&   $W^{\mp}$  &  $Z$   &$\phi_z$ \\ \hline
$WZ\phi\phi_z$&  $\gamma$    & Fig.~26   &$\phi_z$&   $Z$  &  $W$   &$\phi$   \\ \cline{2-7}  
        &  $Z$          & Fig.~29   &$\phi_z$&   $W$  &  $Z$   &$\phi$  \\ \cline{2-7}
        &  $W^{\pm}$        & Fig.~25   &$\phi_z$&   $W^{\mp}$  &  $W^{\mp}$   &$\phi^{\pm}$  \\ \hline
$W\gamma\phi\phi_z$&  $\gamma$    & Fig.~26   &$\phi_z$&   $\gamma$  &  $W$   &$\phi$   \\ \cline{2-7}  
        &  $Z$          & Fig.~29   &$\phi_z$&   $W$  &  $\gamma$   &$\phi$  \\ \hline
$WW\phi\phi$& $\gamma,\ Z$   & Fig.~30   &$\phi$  &   $W$  &  $W$   &$\phi$  \\ \hline
$ZZ\phi\phi$& $\gamma,\ Z$   & Fig.~27   &$\phi$  &   $Z$  &  $Z$   &$\phi$  \\ \cline{2-7}
        &  $W$          & Fig.~28   &$\phi$  &   $W$  &  $Z$   &$\phi$  \\ \hline
$\gamma\gamma\phi\phi$&  $\gamma,\ Z$   & Fig.~27   &$\phi$  &   $\gamma$  &  $\gamma$   &$\phi$  \\ \cline{2-7}
       &   $W$          & Fig.~29   &$\phi$  &   $\gamma$  &  $W$   &$\phi$  \\ \hline
$\gamma Z\phi\phi$& $\gamma,\ Z$   & Fig.~27   &$\phi$  &   $\gamma$  &  $Z$   &$\phi$  \\ \hline
\end{tabular}
\end{center}

\item gSS vertices:  They have already been
discussed in Fig.~15, Fig.~16, and Fig.~17.

\begin{figure}
\vskip -0 cm
\centerline{\epsfxsize 4.7 truein \epsfbox {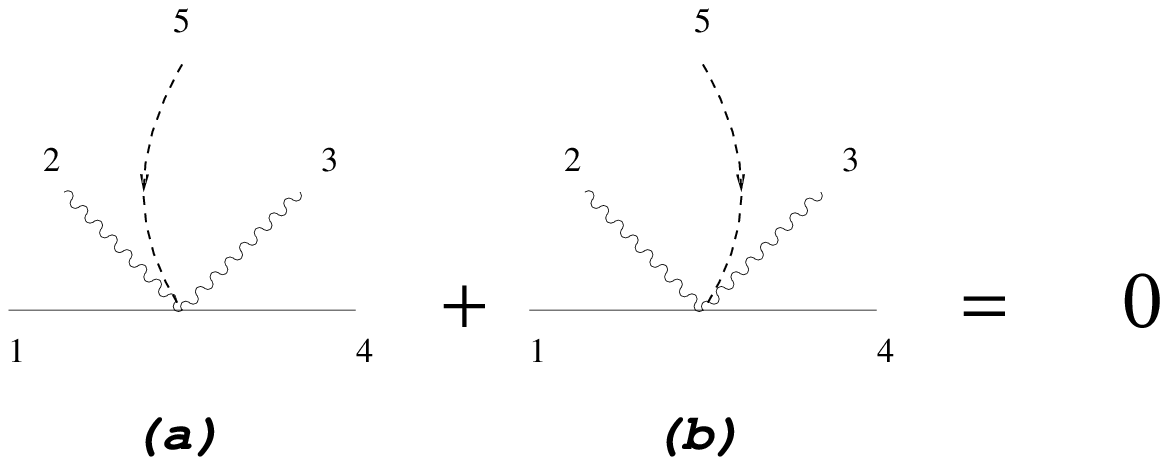}}
\nobreak
\vskip -12.5 cm\nobreak
\vskip .1 cm
\caption{ Cancellation relation of the $ggSS$ vertex.}
\end{figure}

\begin{figure}
\vskip -0 cm
\centerline{\epsfxsize 4.7 truein \epsfbox {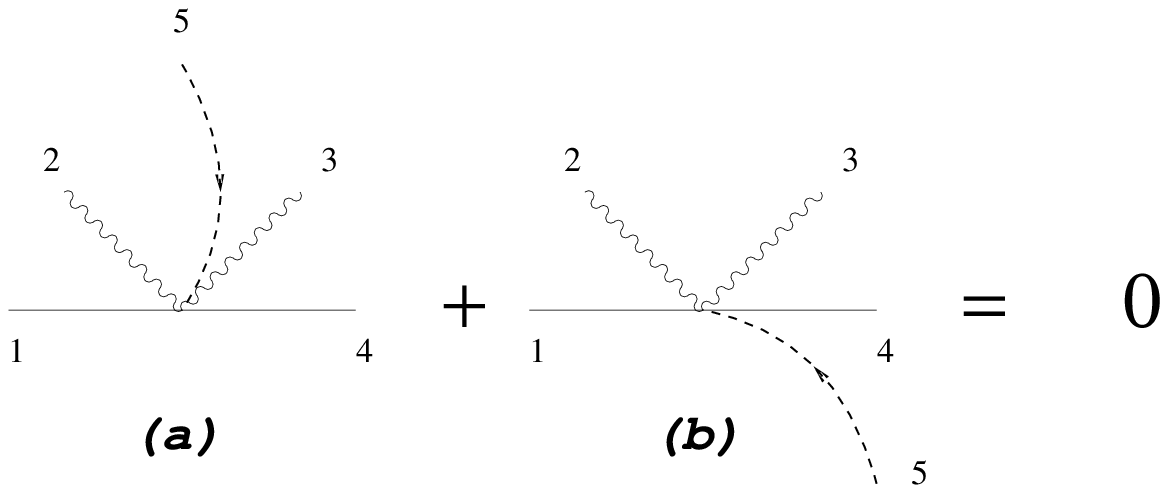}}
\nobreak
\vskip -12.5 cm\nobreak
\vskip .1 cm
\caption{ Cancellation relation of the $ggSS$ vertex.}
\end{figure}

\begin{figure}
\vskip -0 cm
\centerline{\epsfxsize 4.7 truein \epsfbox {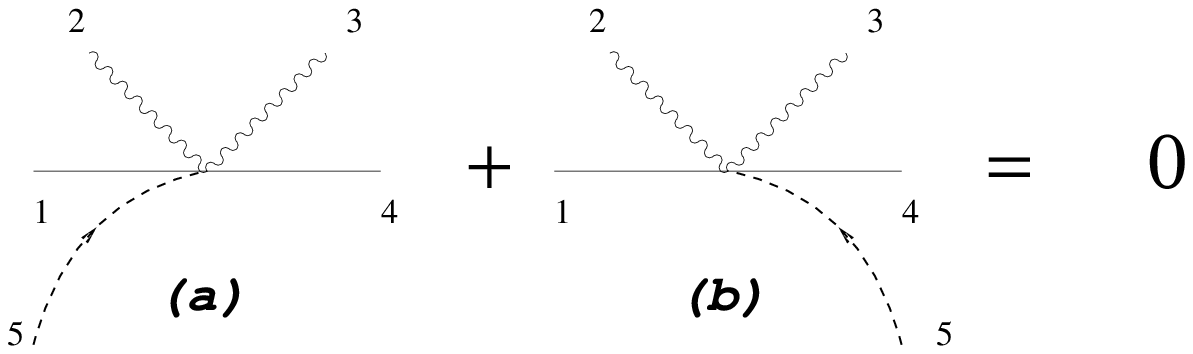}}
\nobreak
\vskip -12.5 cm\nobreak
\vskip .1 cm
\caption{ Cancellation relation of the $ggSS$ vertex.}
\end{figure}

\begin{figure}
\vskip -0 cm
\centerline{\epsfxsize 4.7 truein \epsfbox {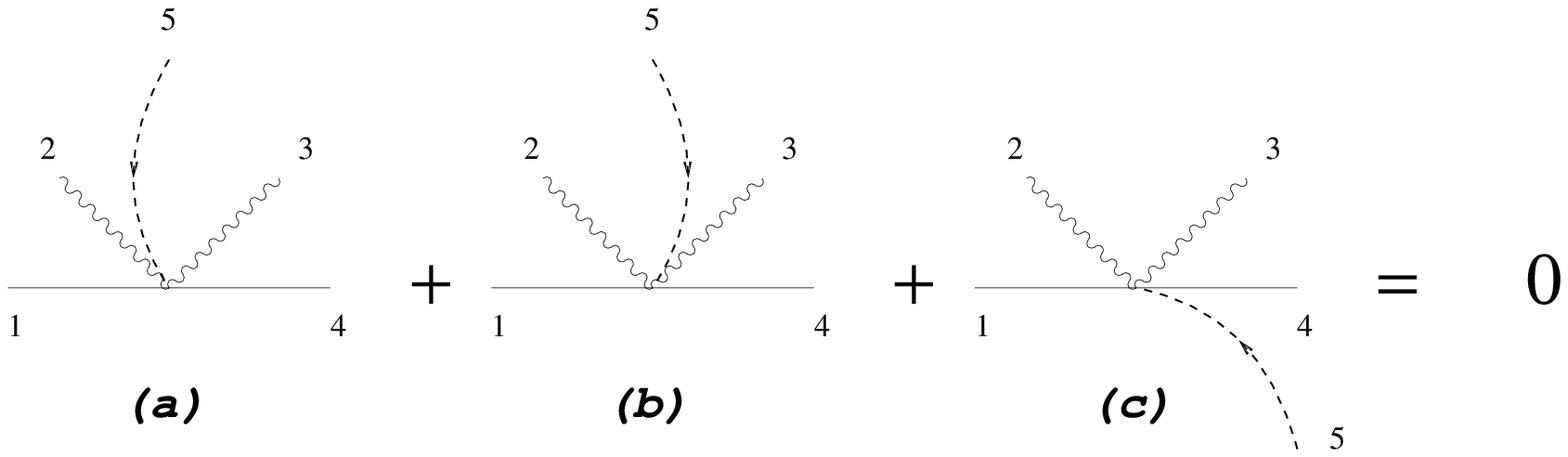}}
\nobreak
\vskip -12.5 cm\nobreak
\vskip .1 cm
\caption{ Cancellation relation of the $ggSS$ vertex.}
\end{figure}

\begin{figure}
\vskip -0 cm
\centerline{\epsfxsize 4.7 truein \epsfbox {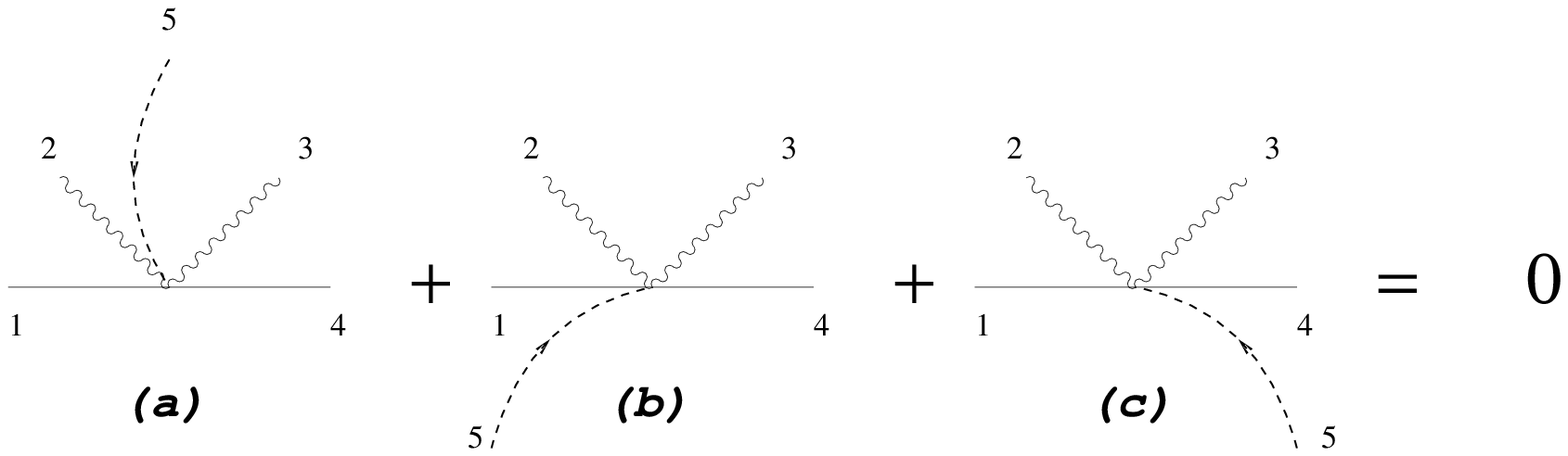}}
\nobreak
\vskip -12.5 cm\nobreak
\vskip .1 cm
\caption{ Cancellation relation of the $ggSS$ vertex.}
\end{figure}

\begin{figure}
\vskip -0 cm
\centerline{\epsfxsize 4.7 truein \epsfbox {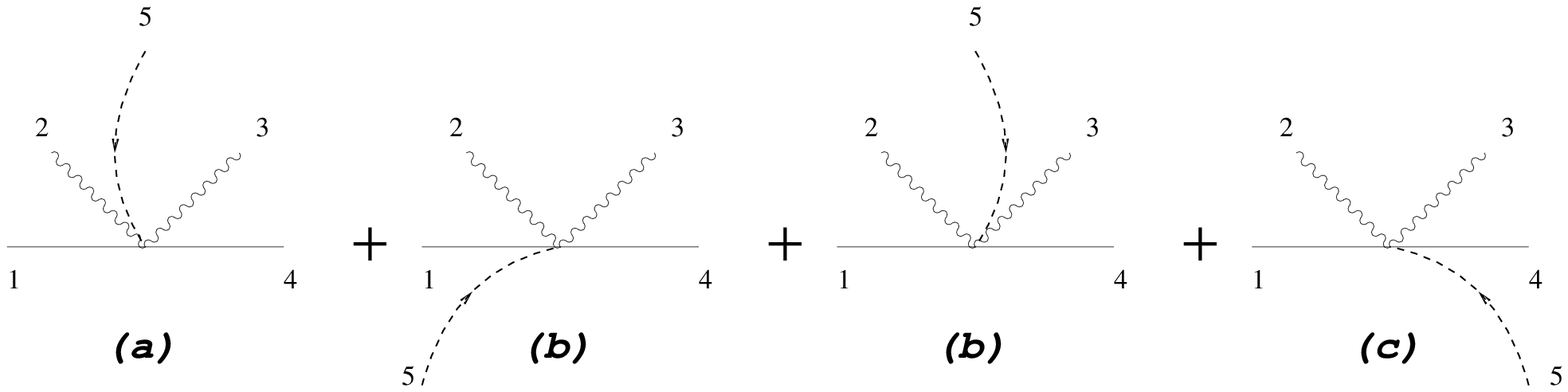}}
\nobreak
\vskip -12.5 cm\nobreak
\vskip .1 cm
\caption{ Cancellation relation of the $ggSS$ vertex.}
\end{figure}

\item gggg:

There are several cases:
\begin{center}
\begin{tabular}{|l|l|l|l|l|l|l|}\hline
vertex   & sliding line & figure & line 1 &line 2 &  line 3  & line 4\\ \hline
$W(\gamma,Z)WWW(\gamma,Z)$ &$\gamma(Z)$  &Fig.~31 & $W^+(\gamma,Z)$ &$W^+$ &$W^-$&$W^-(\gamma,Z)$ 
\\ \hline 
$WWWW(WW\gamma(Z)\gamma(Z))$& $W^{\pm}$&Fig.~32&$\gamma(Z)$&$W^{\mp}$&$W^{\mp}$&$W^{\pm}$ 
\\ \hline
\end{tabular}
\end{center}
\begin{figure}
\vskip -1 cm
\centerline{\epsfxsize 4.7 truein \epsfbox {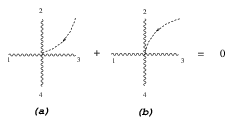}}
\nobreak
\vskip -18.5 cm\nobreak
\vskip .1 cm
\caption{The cancellation of the $gggg$ vertex.}
\end{figure}

\begin{figure}
\vskip -2.5 cm
\centerline{\epsfxsize 4.7 truein \epsfbox {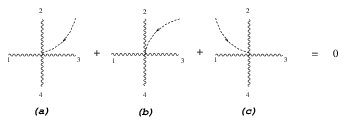}}
\nobreak
\vskip -17.5 cm\nobreak
\vskip .1 cm
\caption{The cancellation of the $gggg$ vertex.}
\end{figure}

\item SSS vertices:
They can all be summarized in the following table:
\begin{center}
\begin{tabular}{|l|l|l|l|l|l|} \hline
vertex       &sliding line  & figure  &   1   &   2   &   3      \\ \hline
$H\phi\phi$  &$\gamma,\ Z$       & Fig.~33   & $\phi$& $\phi$& $H$  \\ \cline{2-6}
             &$Z$           & Fig.~35   & $\phi_z$ &$\phi_z$&$\phi_z$ \\ \cline{2-6}
             &$W^{\pm}$         & Fig.~34   & $\phi^{\mp}$& $\phi^{\mp}$& $\phi^{\pm}$ \\ \hline
$H\phi_z\phi_z$&$W$         & Fig.~33   & $\phi_z$&$\phi$&$H$ \\ \cline{2-6}
             &$W$           & Fig.~35   & $\phi_z$& $\phi$ & $\phi_z$ \\ \cline{2-6}
             &$Z$           & Fig.~35   & $\phi_z$&$\phi_z$&$\phi_z$ \\ \hline
$HHH$        &$W$           & Fig.~35   & $H$    & $\phi$   & $H$   \\ \cline{2-6}
             &$Z$           & Fig.~35   &$H$     & $\phi_z$ & $H$  \\ \hline
\end{tabular}
\end{center}
   
\begin{figure}
\vskip -0 cm
\centerline{\epsfxsize 4.7 truein \epsfbox {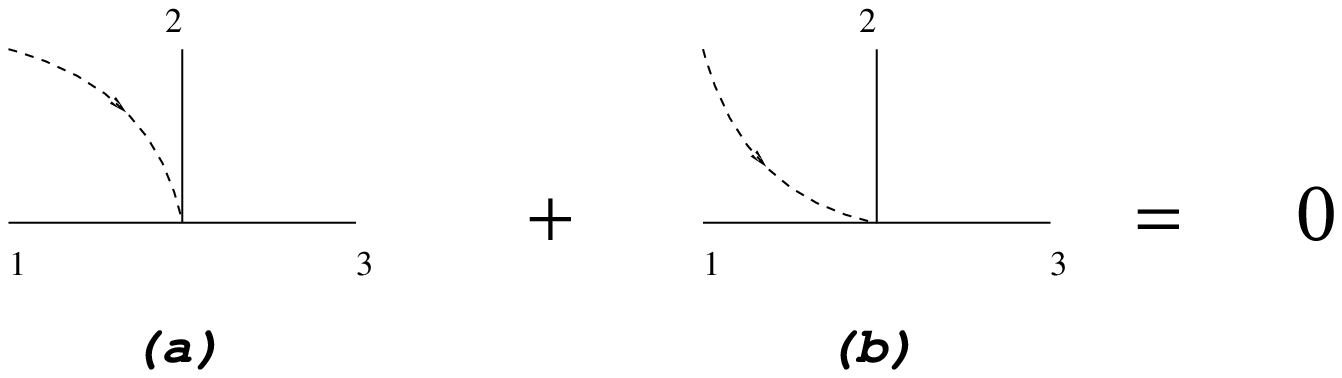}}
\nobreak
\vskip -12.5 cm\nobreak
\vskip .1 cm
\caption{ Cancellation relation of the $SSS$ vertex.}
\end{figure}

\begin{figure}
\vskip -0 cm
\centerline{\epsfxsize 4.7 truein \epsfbox {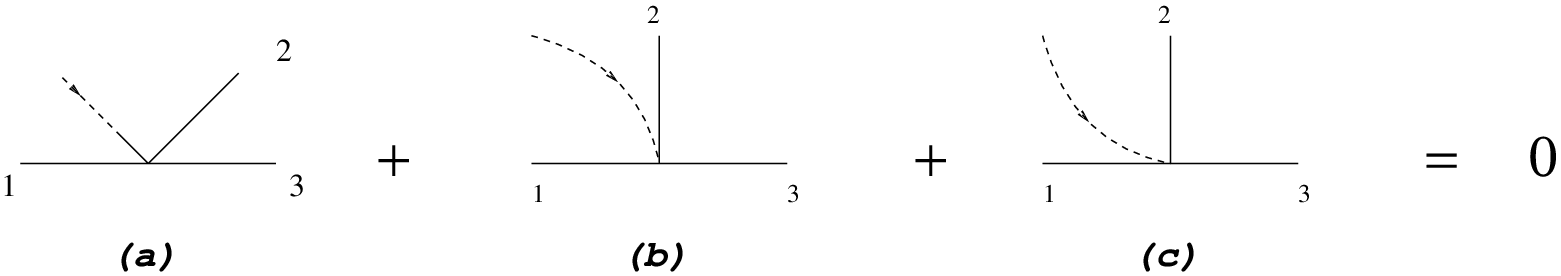}}
\nobreak
\vskip -12.5 cm\nobreak
\vskip .1 cm
\caption{ Cancellation relation of the $SSS$ vertex.}
\end{figure}

\begin{figure}
\vskip -1.5 cm
\centerline{\epsfxsize 4.7 truein \epsfbox {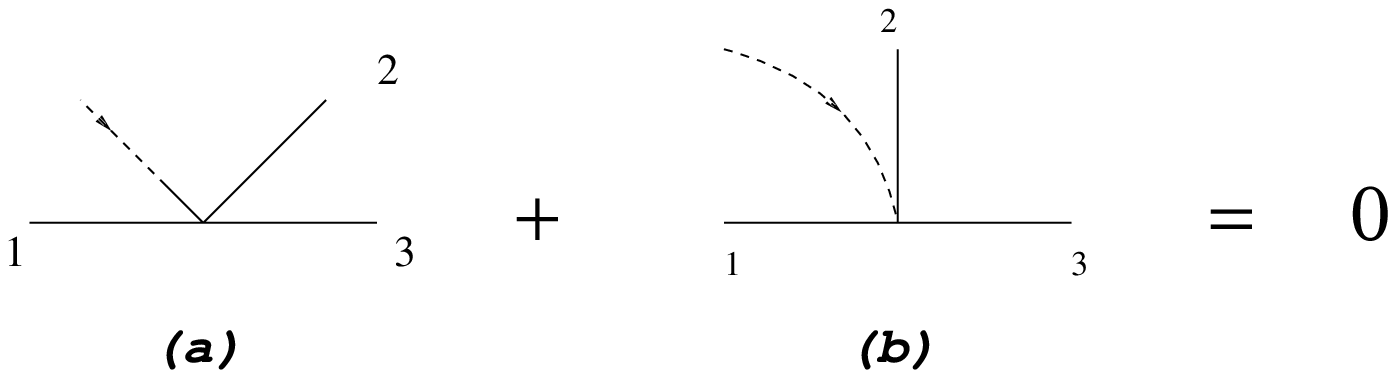}}
\nobreak
\vskip -13 cm\nobreak
\vskip .1 cm
\caption{ Cancellation relation of the $SSS$ vertex.}
\end{figure}

\item SSSS vertices:
\begin{center}
\begin{tabular}{|l|l|l|l|l|l|l|} \hline
vertex   & sliding line& figure & 1   &2   &   3   &  4   \\  \hline
$\phi_z\phi_z\phi_z\phi_z$&$Z$& Fig.~37 & $\phi_z$ &$\phi_z$&$\phi_z$&$H$ \\ \cline{2-7}
     &$W$ &Fig.~37& $\phi_z$&$\phi_z$&$\phi_z$&$\phi$ \\ \hline
$HHHH$ & $Z$ & Fig.~37 & $\phi_z$&$H$&$H$&$H$ \\ \cline{2-7}
       & $W$ &Fig.~35& $\phi$&$H$&$H$&$H$ \\ \hline
$\phi\phi\phi_z\phi_z$& $\gamma,\ Z$  & Fig.~36 & $\phi$&$\phi_z$&$\phi_z$&$\phi$ \\ \cline{2-7}
       & $W$ &Fig.~35 &$\phi$ &$\phi_z$ &$\phi$ &$\phi$ \\ \hline
$HH\phi\phi$&$\gamma,\ Z$& Fig.~35 &$\phi$&$H$&$H$&$\phi$ \\ \cline{2-7}
       & $W^{\pm}$ & Fig.~36 &$\phi^{\mp}$&$\phi^{\pm}$&$H$&$\phi^{\mp}$  \\ \cline{2-7}
       & $Z$   &Fig.~35&$\phi$&$H$&$\phi_z$&$\phi$  \\ \hline
$HH\phi_z\phi_z$ & $W$&Fig.~37                                                                                                         &$H$&$H$&$\phi$&$\phi_z$ \\ \hline
$\phi\phi\phi\phi$ & $\gamma, Z$ & Fig.~35& $\phi$&$\phi$&$\phi$&$\phi$ \\ \hline
\end{tabular}
\end{center}

\begin{figure}
\vskip -1 cm
\centerline{\epsfxsize 4.7 truein \epsfbox {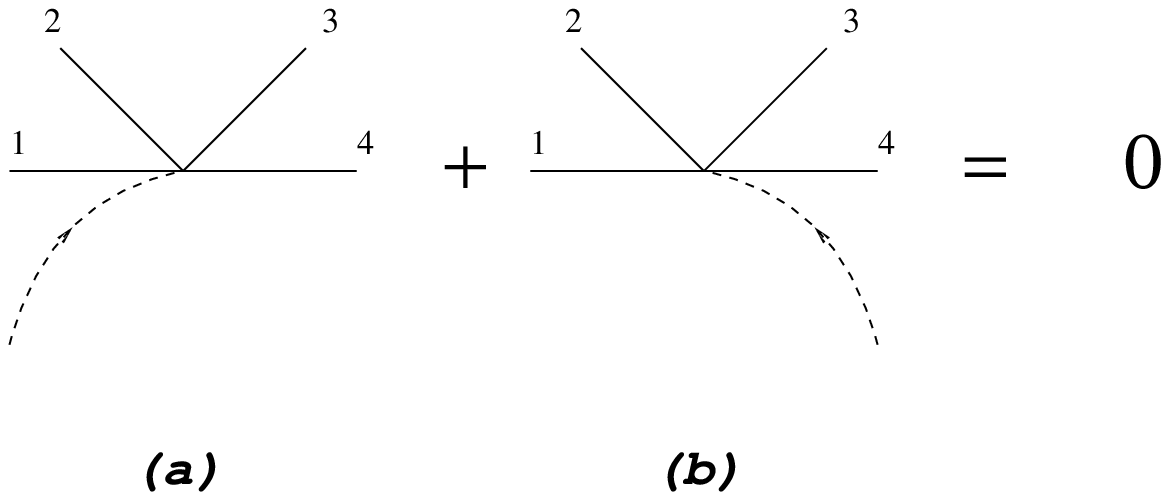}}
\nobreak
\vskip -12.5 cm\nobreak
\vskip .1 cm
\caption{ The cancellation relation of the $SSSS$ vertex.}
\end{figure}

\begin{figure}
\vskip -0 cm
\centerline{\epsfxsize 4.7 truein \epsfbox {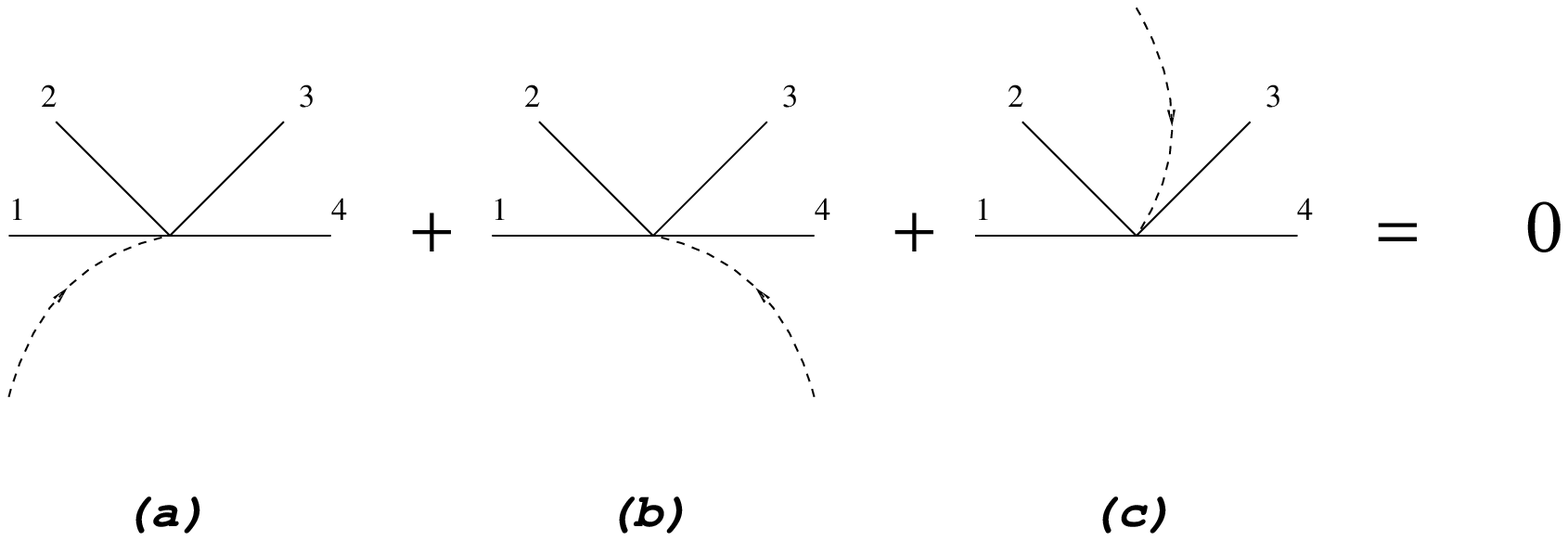}}
\nobreak
\vskip -12.5 cm\nobreak
\vskip .1 cm
\caption{The cancellation relation of the $SSSS$ vertex. }
\end{figure}

\begin{figure}
\vskip -0 cm
\centerline{\epsfxsize 4.7 truein \epsfbox {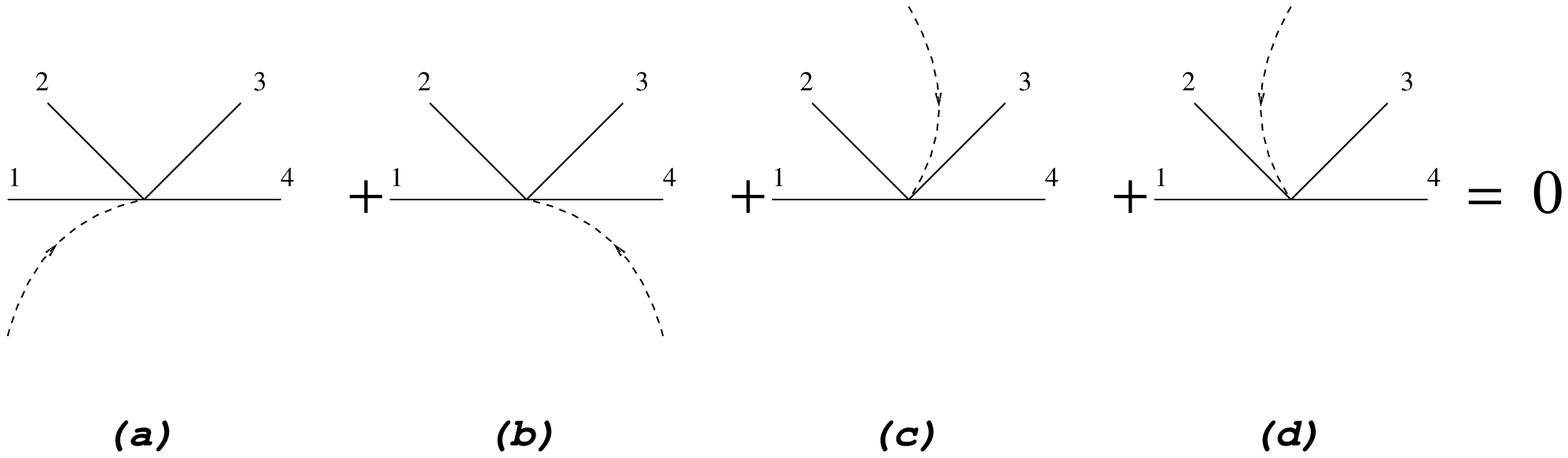}}
\nobreak
\vskip -11.5 cm\nobreak
\vskip .1 cm
\caption{ The cancellation relation of the $SSSS$ vertex.}
\end{figure}

\item gGG:  
\begin{center}
\begin{tabular}{|l|l|l|l|l|l|}\hline
vertex     & figure        &  1          &  2    & 3    &  4  \\ \hline
$\bar{c}^{\pm} \gamma(Z)c^{\mp}$ &Fig.~39 & $\bar{c^{\pm}}$   &$ W^{\mp}$   &  $c^+ $&$ c^-$ \\ \hline
$\bar{c}^{\pm} \gamma(Z)c^{\mp}$ &Fig.~40 & $\bar{c^{\pm}}$   &$ W^{\mp}$   & $ c^{\mp}$ & $c^{\mp}$ \\ \hline
$\bar{c}^{\pm} W^{\mp} c_\gamma(c_z)$&Fig.~40 & $\bar{c^{\pm}}$& $W^{\mp}$  &$c_\gamma(c_z)$ 
& $c_\gamma(c_z)$ \\ \hline
$\bar{c}_{\gamma}(\bar{c}_z)W^{\pm}c^{\mp}$&Fig.~39&
$\bar{c}_{\gamma}(\bar{c}_z)$ &$W^{\pm}$&$c_\gamma(c_z)$ &$c^{\mp}$ \\ \hline
\end{tabular}
\end{center}

\begin{figure}
\vskip -1.5 cm
\centerline{\epsfxsize 4.7 truein \epsfbox {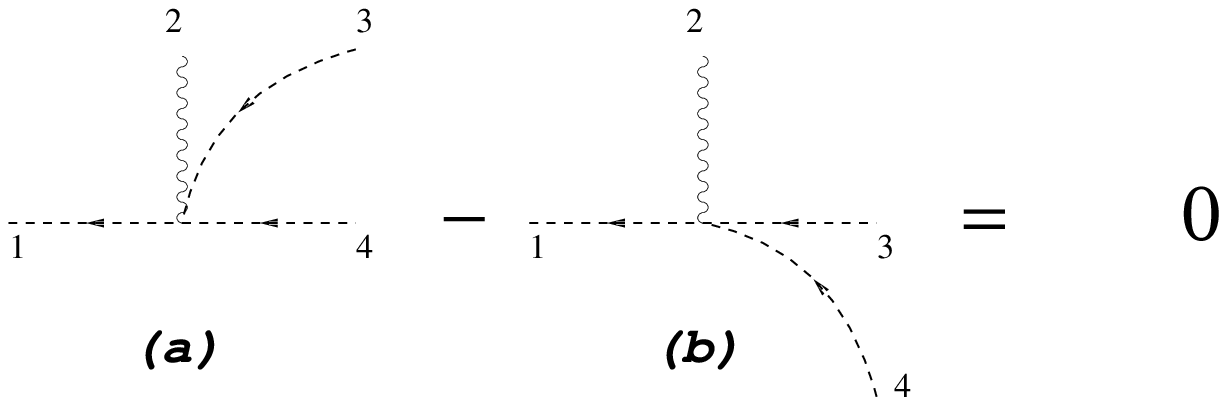}}
\nobreak
\vskip -13 cm\nobreak
\vskip .1 cm
\caption{ The cancellation relation of the $gGG$ vertex.}
\end{figure}

\begin{figure}
\vskip -.5 cm
\centerline{\epsfxsize 4.7 truein \epsfbox {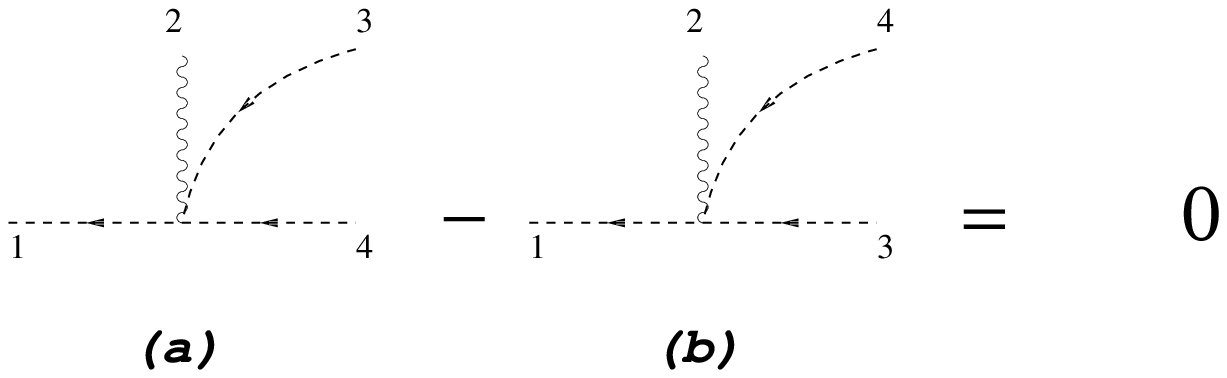}}
\nobreak
\vskip -12.5 cm\nobreak
\vskip .1 cm
\caption{ The cancellation relation of the $gGG$ vertex.}
\end{figure}

\item SGG:
All the cancellations of this kind of vertices is represented by Fig.~41.
\begin{figure}
\vskip -0 cm
\centerline{\epsfxsize 4.7 truein \epsfbox {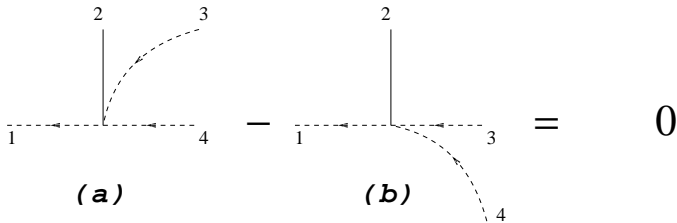}}
\nobreak
\vskip -12.5 cm\nobreak
\vskip .1 cm
\caption{ The cancellation relation of the $SGG$ vertex.}
\end{figure}
\end{enumerate}

\subsection{Cancellation about the mass compensation diagrams}
All of these have been considered in the above discussion.


\section{GN gauge}
\subsection{Divergence relations}
First we look at the divergence relations of the vertices in the GN gauge. 
Since most 
of them are the same as those in the ordinary gauge, we just need to discuss
 those that are different. 
\begin{enumerate}
\item ggg:
It is shown in Fig.~42.
\begin{figure}
\vskip -0 cm
\centerline{\epsfxsize 4.7 truein \epsfbox {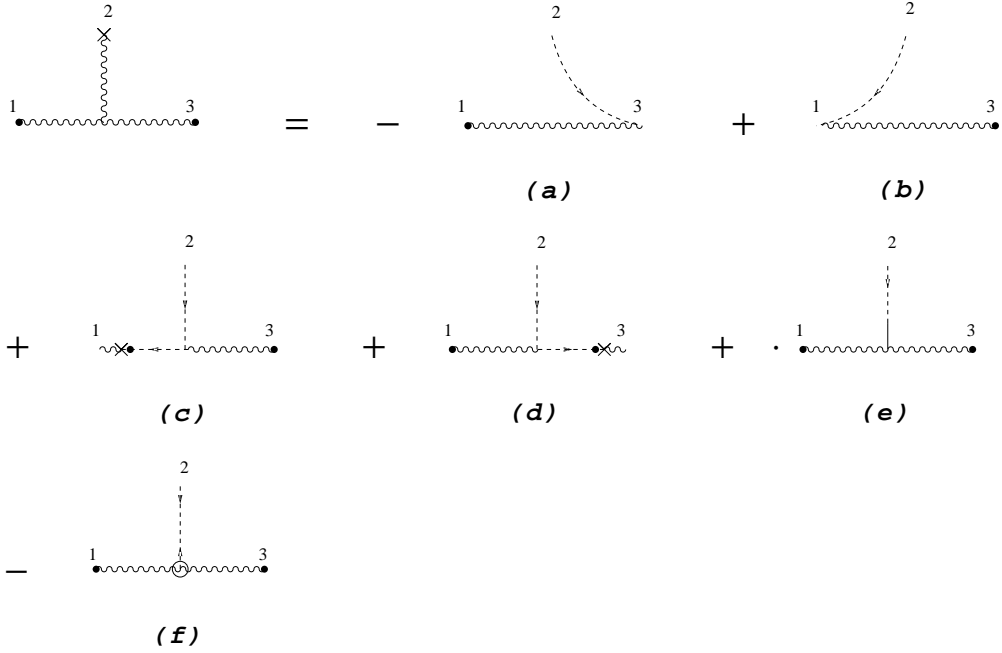}}
\nobreak
\vskip -8.5cm\nobreak
\vskip .1cm
\caption{divergence relation of the triple gauge boson vertex in GN gauge.}
\end{figure}

Here the extra diagram (f) is called {\it stagnant diagram} with a ghost
line that can go nowhere. 

\item gggg:
We use the following table to summarize several cases:

\begin{center}
\begin{tabular}{|l|l|l|l|l|l|}\hline
Vertex     & figure      & 1      & 2       & 3      & 4    \\ \hline
$WWWW$       & Fig.~43     & $W$      & $W$      & $W$      & $W$    \\ \hline
$WW\gamma(Z)\gamma(Z)$ & Fig.~43    & $W$      & $\gamma(Z)$    & $W$      & $\gamma(Z)$ \\ \hline
$WW\gamma(Z)\gamma(Z)$ & Fig.~44     & $\gamma(Z)$   & $W$       & $\gamma(Z)$   & $W$    \\ \hline
$W\gamma WZ$       & Fig.~44     & $\gamma(Z)$   & $W$       & $Z(\gamma)$   & $W$    \\ \hline
$W\gamma WZ$       & Fig.~43     & $W $     & $\gamma$       & $W$      & $Z$    \\ \hline
\end{tabular}
\end{center}

\begin{figure}
\vskip -0 cm
\centerline{\epsfxsize 4.7 truein \epsfbox {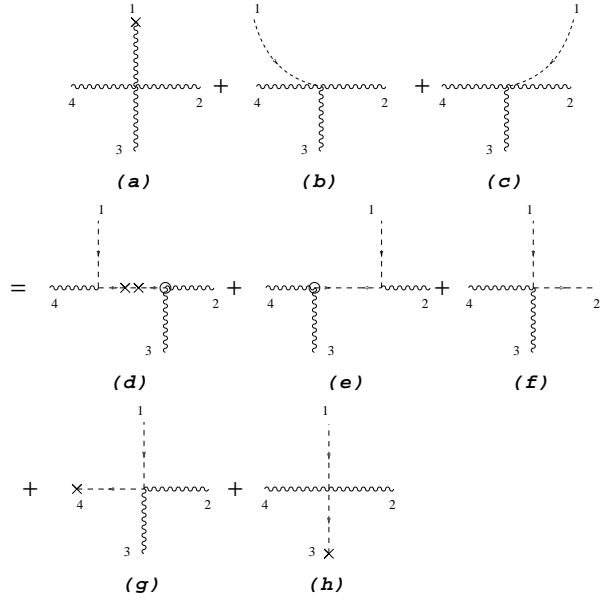}}
\nobreak
\vskip -6.5cm\nobreak
\vskip .1cm
\caption{The divergence relation of the four gauge boson vertex in the 
GN gauge.}
\end{figure}

\begin{figure}
\vskip -0 cm
\centerline{\epsfxsize 4.7 truein \epsfbox {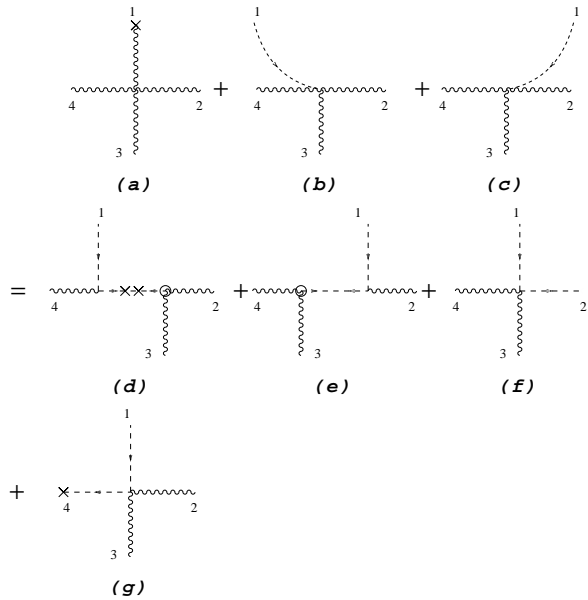}}
\nobreak
\vskip -8.5cm\nobreak
\vskip .1cm
\caption{The divergence relation of the $4g$-vertex in the GN gauge.}
\end{figure}

\item gGG:

The divergence relation depends on whether the ghost vertex is a-type or 
b-type. The b-type ghost vertices have similar divergence relation as
the charged scalar. 
\begin{center}
\begin{tabular}{|l|l|l|l|l|}\hline
vertex           & figure     & 1     & 2    & 3                   \\ \hline
$\bar{c}^-W^+c_\gamma(c_z)$  & Fig.~45    & $\bar{c^-}$ & $W^+$   & $c_\gamma(c_z)$  \\ 
\hline
$\bar{c}^+\gamma(Z) c^-$     & Fig.~46    & $\bar{c^+}$ & $\gamma(Z)$  & $c^-$       \\
 \hline
$\bar{c}_\gamma(\bar{c}_z)W^-c^+$ &Fig.~45 &$\bar{c}_\gamma(\bar{c}_z)$&$W^-$&$c^+$ \\ 
\hline
\end{tabular}
\end{center}
\begin{figure}
\vskip -.5 cm
\centerline{\epsfxsize 4.7 truein \epsfbox {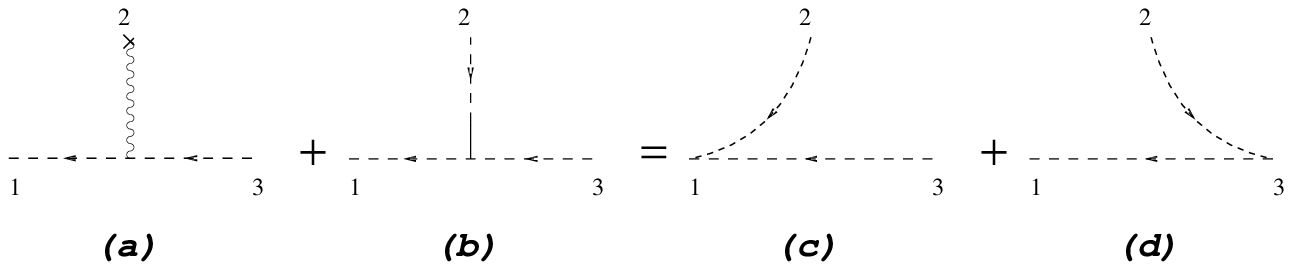}}
\nobreak
\vskip -13.5cm\nobreak
\vskip .1cm
\caption{The divergence relation of the b-type ghost  vertex in GN gauge.}
\end{figure}

\begin{figure}
\vskip -1 cm
\centerline{\epsfxsize 4.7 truein \epsfbox {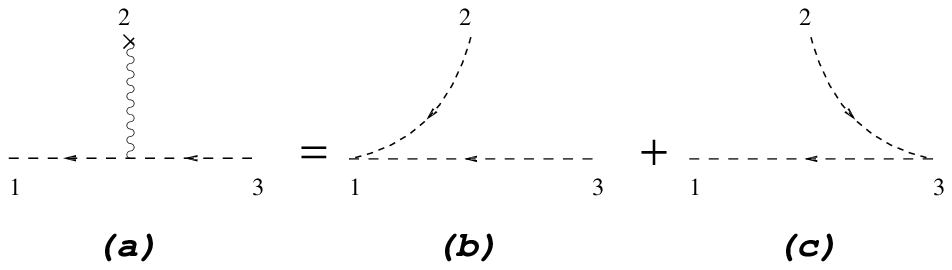}}
\nobreak
\vskip -13.5cm\nobreak
\vskip .1cm
\caption{The divergence relation of the b-type ghost vertex in the GN gauge.}
\end{figure}

As for the a-type ghost vertices, they are cancelled according to the following
table:

\begin{center}
\begin{tabular}{|l|l|l|l|l|l|l|}\hline
vertex  & figure   & line1  &line2 &line3 &line4 &$\lambda$\\ \hline
$W^-\bar{c}^+ c_\gamma(c_z)$  & Fig.~47  &$\bar{c}^+$ &$\gamma(Z)$ & $c^-$  &$c_\gamma(c_z)$
& 1 
\\ \cline{2-7}
 &Fig.~47 &$\bar{c}^+$ &$W^-$ & $c_\gamma(c_z)$  &$c_\gamma(c_z)$ &1
\\ \hline
$\bar{c}_\gamma(\bar{c}_z)W^+c^- $ & Fig.~47 &$\bar{c}_\gamma(\bar{c}_z) $&$W^+$&$c_\gamma(c_z)$&
$c^-$ &1\\ \cline{2-7}
 &Fig.~47 &$\bar{c}_\gamma(\bar{c}_z) $&$\gamma(Z)$&$c^+$&
$c^-$&1\\ \hline
$\bar{c}^-\gamma(Z)c^+$& Fig.~47 & $\bar{c}^-$ & $W^-$& $c^+$ & $c^+$&0 \\ 
\cline{2-7} 
 &Fig.~47 & $\bar{c}^-$ & $W^+$& $c^-$ & $c^+$&0 \\ 
\hline 
\end{tabular}
\end{center}
\begin{figure}
\vskip -1 cm
\centerline{\epsfxsize 4.7 truein \epsfbox {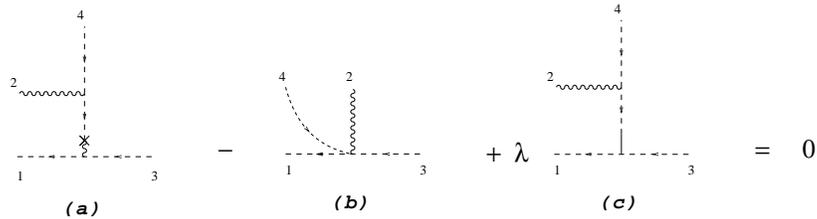}}
\nobreak
\vskip -13.5cm\nobreak
\vskip .1cm
\caption{The divergence relation of the a-type ghost vertex.}
\end{figure}

\item ggS:
All the identities are the same as in Feynman gauge graphically, although 
now the vertices become those in GN gauge.

\end{enumerate}

\subsection{Cancellation relations}
\begin{enumerate}
\item ggg:

They are contained in the discussion of the divergence relation of gggg above.

\item gggg:

They are the same as in Feynman gauge, Although the gggg vertex has different
expression.

\item gGG:

All the possibilities of a ghost line being sliding into a gGG vertex 
are classified into the table below.
\begin{center}
\begin{tabular}{|l|l|l|l|l|}\hline
figure  &line 1          & line 2   & line 3 & line 4 \\ \hline
Fig.~48 &$\bar{c}^{\pm}$ &$c^{\mp}$ &$\gamma(Z) $ &$c_\gamma(c_z)$ \\ \cline{2-5}
        &$\bar{c}^{\pm}$ &$c^{\pm}$ &$c^{\mp}$&$W^{\mp}$ \\ \hline
Fig.~49 &$\bar{c}^{\pm}$ &$c_\gamma(c_z)$&$c_\gamma(c_z)$&$W^{mp}$ \\ \cline{2-5}
        &$\bar{c}^{\pm}$ &$c^{\mp}$&$c^{\mp}$&$W^{\pm}$ \\ \cline{2-5}
        &$\bar{c}_\gamma(\bar{c}_z)$&$c^+$&$c^-$ &$\gamma(Z)$  \\ \hline
Fig.~50 &$\bar{c}_\gamma(\bar{c}_z)$&$c_\gamma(c_z)$&$c^{\pm}$ &$W^{\mp}$  \\ \hline
\end{tabular}
\end{center}
\begin{figure}
\vskip -2 cm
\centerline{\epsfxsize 4.7 truein \epsfbox {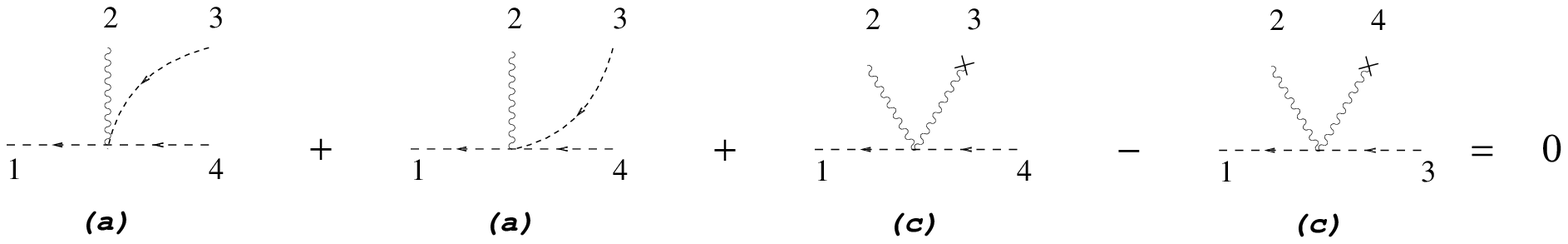}}
\nobreak
\vskip -12.5cm\nobreak
\vskip .1cm
\caption{The cancellation relation of the $gGG$ vertex in the GN gauge.}
\end{figure}
\begin{figure}
\vskip -1 cm
\centerline{\epsfxsize 4.7 truein \epsfbox {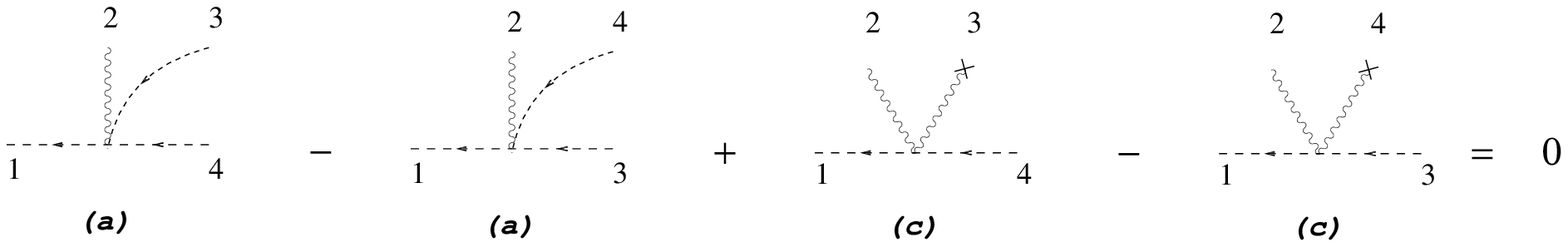}}
\nobreak
\vskip -12.5cm\nobreak
\vskip .1cm
\caption{The cancellation relation of the $gGG$ vertex in the GN gauge.}
\end{figure}
\begin{figure}
\vskip -5 cm
\centerline{\epsfxsize 4.7 truein \epsfbox {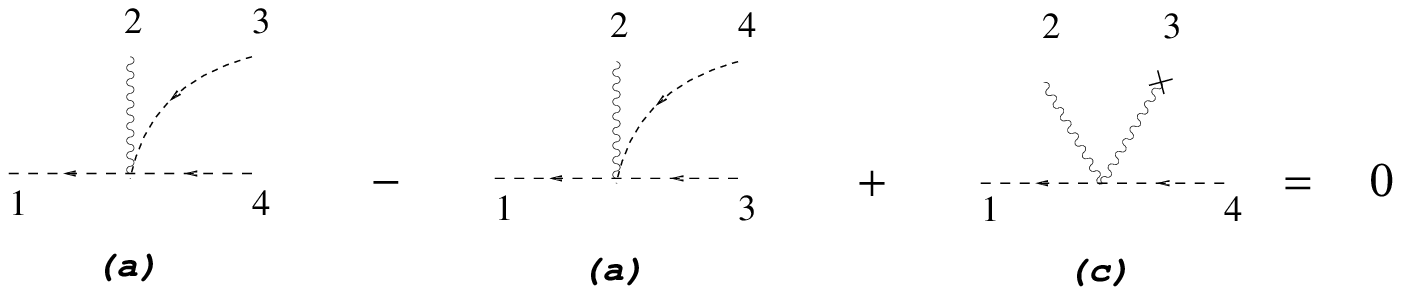}}
\nobreak
\vskip -12.5cm\nobreak
\vskip .1cm
\caption{The cancellation relation of the $gGG$ vertex in the GN gauge.}
\end{figure}

\item ggS:

They are 
the same as in Feynman gauge, although the vertex factors are now different.

\item ggGG:

Since there are many possibilities here, we classify them in a new way. 
First if a ghost line slide
to a ggGG vertex, we are then looking at a configuration with one 
outgoing ghost, two incoming ghosts,
and two bosons. If those two ghost lines are the same, then any of 
them can be the wandering ghost line.
These two cases just cancel each other. So in the following we just 
focus on the cases with different
incoming ghost lines.
\begin{center}
\begin{tabular}{|l|l|l|l|l|l|}\hline
line 1& line 2 & line 3 & line 4   & line 5 & figure  \\ \hline
$\bar{c}^+$&$c^-$&$c_\gamma(c_z)$&$\gamma(Z)$&$\gamma(Z)$&Fig.~51 \\ \hline
$\bar{c}^-$&$c^+$&$c_\gamma(c_z)$&$\gamma(Z)$&$\gamma(Z)$&Fig.~52 \\ \hline
$\bar{c}^+$&$c^+$&$c^-$&$W^-$&$\gamma(Z)$&Fig.~53      \\ \hline
$\bar{c}^-$&$c^-$&$c^+$&$W^+$&$\gamma(Z)$&Fig.~51     \\ \hline
$\bar{c}_\gamma(\bar{c}_z$&$c_\gamma(c_z)$&$c^-$&$W^+$&$\gamma(Z)$&Fig.~54 \\ \hline
$\bar{c}_\gamma(\bar{c}_z$&$c_\gamma(c_z)$&$c^+$&$W^-$&$\gamma(Z)$&Fig.~55 \\ \hline
$\bar{c}_\gamma(\bar{c}_z$&$c^+$&$c^-$&$\gamma(Z)$&$\gamma(Z)$&Fig.~56 \\ \hline
\end{tabular}
\end{center}
\begin{figure}
\vskip -1 cm
\centerline{\epsfxsize 4.7 truein \epsfbox {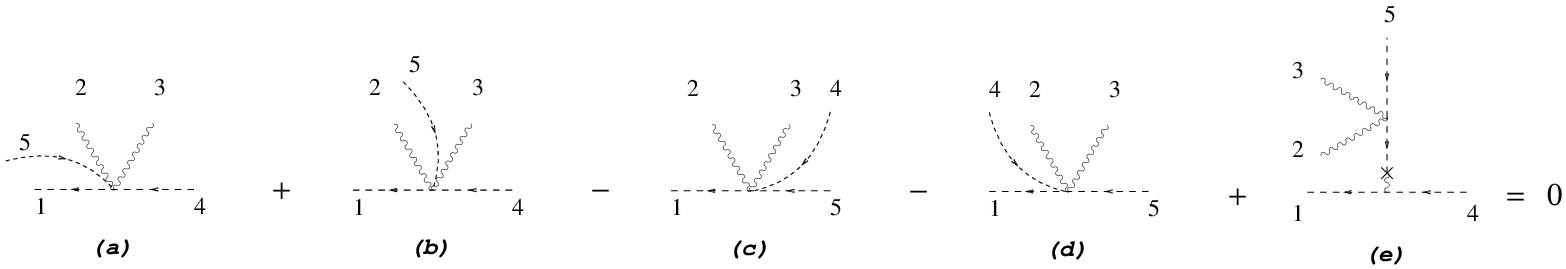}}
\nobreak
\vskip -13.5cm\nobreak
\vskip .1cm
\caption{The cancellation relation of the $ggGG$ vertex in the GN gauge.}
\end{figure}

\begin{figure}
\vskip -1 cm
\centerline{\epsfxsize 4.7 truein \epsfbox {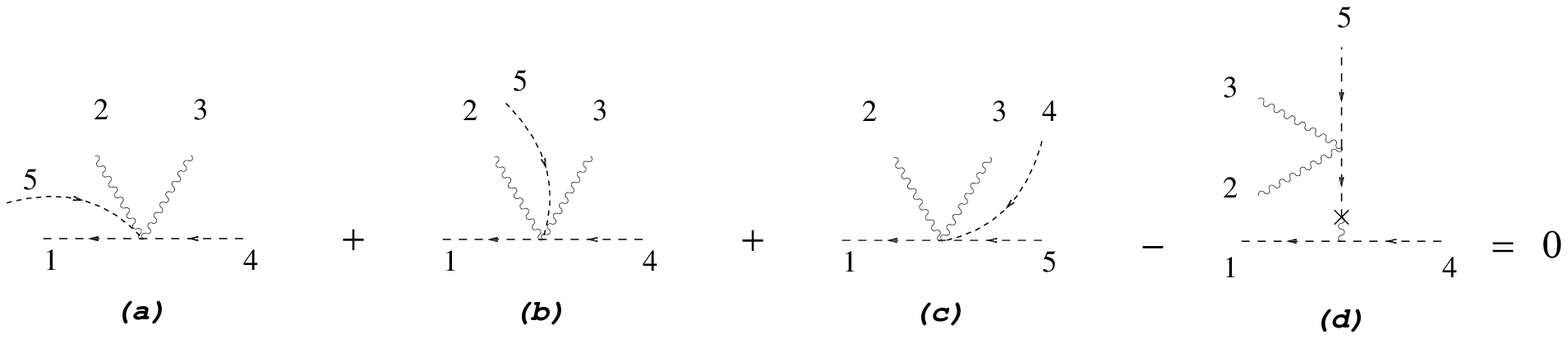}}
\nobreak
\vskip -12.5cm\nobreak
\vskip .1cm
\caption{The cancellation relation of the $ggGG$ vertex in the GN gauge.}
\end{figure}
\begin{figure}
\vskip -1 cm
\centerline{\epsfxsize 4.7 truein \epsfbox {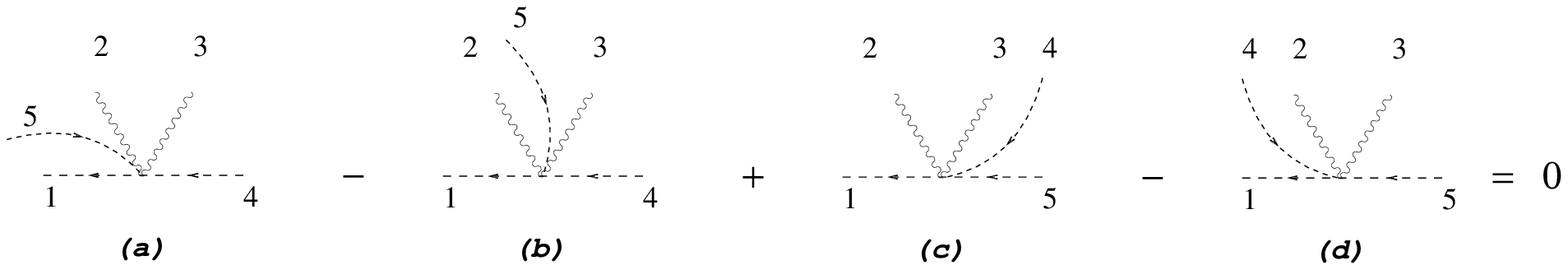}}
\nobreak
\vskip -12.5cm\nobreak
\vskip .1cm
\caption{The cancellation relation of the $ggGG$ vertex in the GN gauge.}
\end{figure}
\begin{figure}
\vskip -0 cm
\centerline{\epsfxsize 4.7 truein \epsfbox {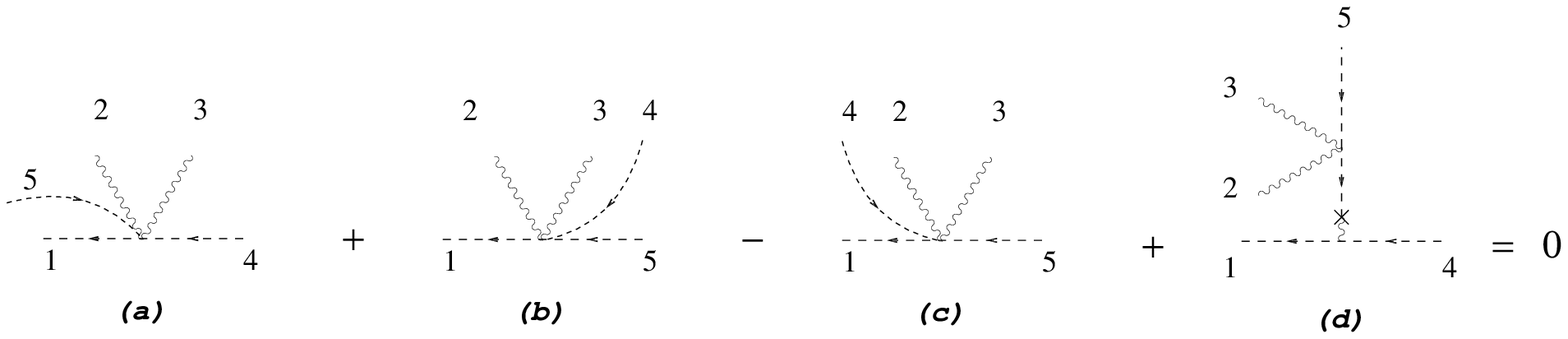}}
\nobreak
\vskip -12.5cm\nobreak
\vskip .1cm
\caption{The cancellation relation of the $ggGG$ vertex in the GN gauge.}
\end{figure}
\begin{figure}
\vskip -1 cm
\centerline{\epsfxsize 4.7 truein \epsfbox {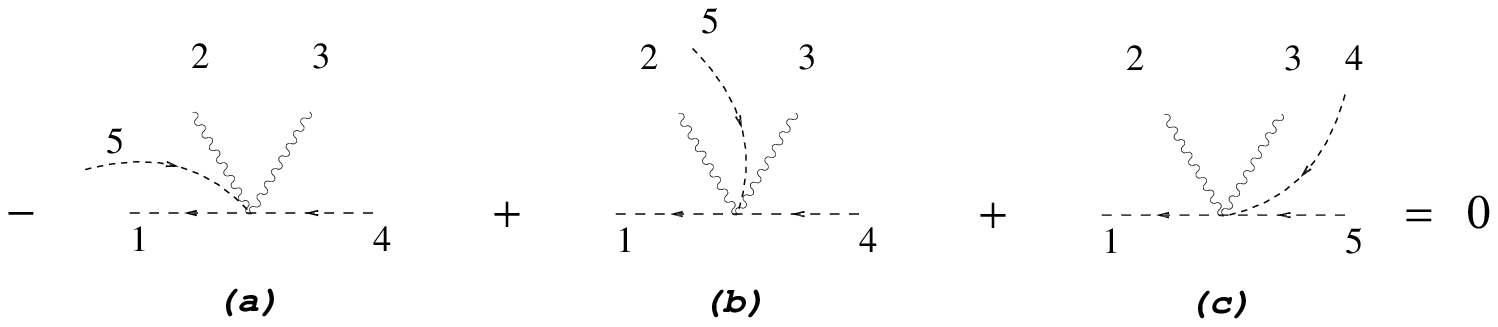}}
\nobreak
\vskip -12.5cm\nobreak
\vskip .1cm
\caption{The cancellation relation of the $ggGG$ vertex in the GN gauge.}
\end{figure}
\begin{figure}
\vskip -0 cm
\centerline{\epsfxsize 4.7 truein \epsfbox {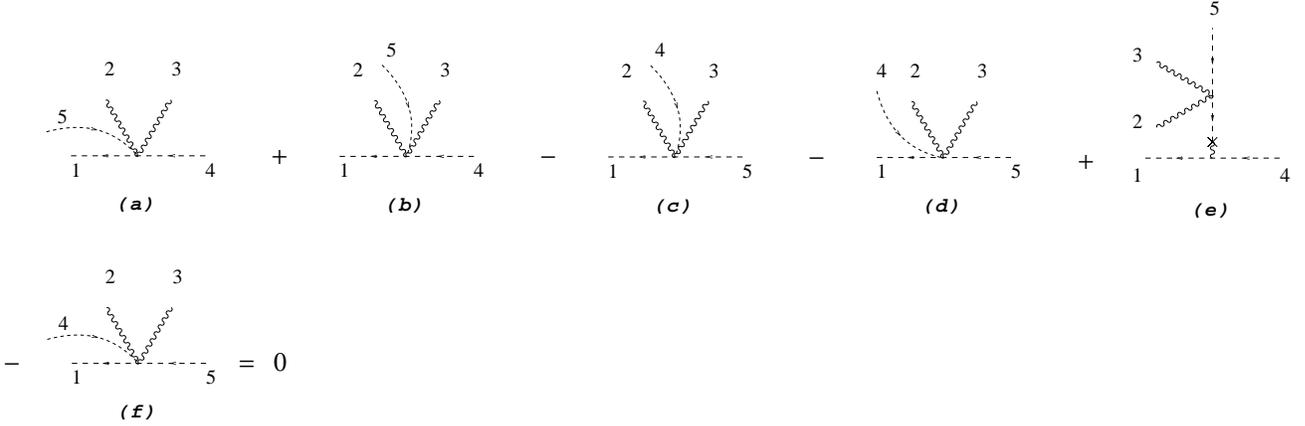}}
\nobreak
\vskip -11.5cm\nobreak
\vskip .1cm
\caption{The cancellation relation of the $ggGG$ vertex in the GN gauge.}
\end{figure}
\end{enumerate}


\newpage
\begin{thebibliography}{99}

\bibitem[*]{YF}e-mail address: feng@physics.mcgill.ca

\bibitem[\dag]{CL}e-mail address: lam@physics.mcgill.ca

\bibitem[1]{YJ} Y.J. Feng and C.S. Lam, Phys. Rev {\bf D53} (1996), 2115.

\bibitem[2]{new} Y.J. Feng and C.S. Lam, to be published.

\bibitem[3]{RG} R. Gastmans, Tai Tsun Wu, {\it The  Ubiquitous Photon: Helicity
 method for QED and QCD}, Clarendon Press (1990).\\
            M.L. Mangano, S.J. Parke, Phys. Rep. {\bf 200}, 301 (1991).

\bibitem[4]{CS} C.S. Lam and J.P. Lebrun, Nuovo Cimento A {\bf 59}, 
               397 (1969).\\
              C.S. Lam, Nucl. Phys. {\bf B397}, 143 (1993).

\bibitem[5]{ZB} Z. Bern and D.C. Dunbar, Nucl. Phys. {\bf B379}, 562
             (1992).\\
Z. Bern and D.A. Kosower, Phys. Rev. Lett. {\bf 66}, 1669 (1991); Nucl. Phys.
{\bf B362}, 389 (1991); {\bf B379}, 451 (1992).\\
Z. Bern, L. Dixon, and D.A. Kosower, Phys. Rev. Lett. {\bf 70}, 2677 (1993).

\bibitem[6]{MS} M. Strassler, Nucl. Phys. {\bf B385}, 145 (1992); SLAC preprint 
              No. SLAC-PUB 5978, 1992 (unpublished).\\
              M.G. Schmidt and C. Schubert, Phys. Lett. B {\bf 331}, 69 (1994).
\bibitem[7]{JL} J.L. Gervais and A. Neveu, Nucl. Phys. {\bf B46}, 381 (1972).

\bibitem[8]{BS}  B.S. DeWitt, Phys. Rev. {\bf 162} (1967) 1195, 1239; in {\it
       Dynamic theory of groups and fields} (Gordon and Breach, 1965).\\
               G. 't Hooft, Nucl. Phys. {\bf B62}, 444 (1973).\\
               L.F. Abbott, Nucl. Phys. {\bf B185}, 189 (1981).

\bibitem[9]{XL} X. Li and Y. Liao, ASITP-94-50, hep-ph 9409401.\\
              A. Denner, G. Weiglein, and S. Dittmaier, hep-ph 9410338,
              Nucl. Phys. {\bf B440}, 95 (1995).

\bibitem[10]{JP} J. Papavassiliou and K. Philippides, hep-ph 9503377;
               Phys. Rev. {\bf D52} (1995), 2355; \\
               J. Papavassiliou and A. Pilaftsis, hep-ph 9506417,
               Phys. Rev. Lett. {\bf 75}, 3060 (1995).

\bibitem[11]{Sterman} G. Sterman, {\it Introduction to Quantum Field Theory}, Cambridge University Press 1993.
\bibitem[12]{JM} J.M. Cornwall, D.N. Levin, and G. Tiktopoulos, Phys. Rev. {\bf D10} (1974), 1145; \\
                  H.J. He, Yu-Ping Kuang, and C.-P. Yuan, YPI-IHEP-95-05,
                  Phys.Rev. {\bf D51} (1995), 6463. 
               
\end{thebibliography}
\end{document}